\documentclass[useAMS,usenatbib]{mn2e}
\usepackage{natbib}
\usepackage{enumerate}
\usepackage{amsmath}
\usepackage{amssymb}
\usepackage{float}
\usepackage{threeparttable}
\usepackage{array}
\usepackage{graphicx}
\usepackage{float}
\usepackage{lscape}
\usepackage{hyperref}
\newcommand{\fitrise}{$t_{\mathrm{rise}}^{\mathrm{end}}$}
\newcommand{\rise}{$t_{\mathrm{rise}}^{\mathrm{max}}$}

\newcommand{\lambdaeff}{$\lambda_{\mathrm{eff}}$}
\newcommand{\rsun}{$R_{\odot}$}
\newcommand{\msun}{$M_{\odot}$}
\newcommand{\tmax}{$t_{\mathrm{max}}^{\mathrm{Baz}}$}
\newcommand{\tend}{$t_{\mathrm{end}}^{\mathrm{Baz}}$}
\newcommand{\texp}{$t_{\mathrm{exp}}^{\mathrm{pow}}$}

\title{The Rise-Time of Type II Supernovae}

\author[Gonz\'alez-Gait\'an et al.]{S. Gonz\'alez-Gait\'an$^{1,2}$\thanks{E-mail: sgonzale@das.uchile.cl}, N. Tominaga$^{3,4}$, J. Molina$^{2}$, L. Galbany$^{1,2}$, F. Bufano$^{1,5}$,
\newauthor
J. P. Anderson$^6$, C. Gutierrez$^{1,2,6}$, F. F\"orster$^{1,7}$, G. Pignata$^{1,5}$, M. Bersten$^{8,9,4}$,
\newauthor
D. A. Howell$^{10,11}$, M. Sullivan$^{12}$, R. Carlberg$^{13}$, T. de Jaeger$^{1,2}$, M. Hamuy$^{2,1}$, 
\newauthor
P. V. Baklanov$^{14,15}$,S. I. Blinnikov$^{14,16,4}$ \\
$^1$ Millennium Institute of Astrophysics, Casilla 36-D, Santiago, Chile \\
$^2$ Departamento de Astronom\'ia, Universidad de Chile, Camino El Observatorio 1515, Las Condes, Santiago, Chile \\
$^3$ Department of Physics, Faculty of Science and Engineering, Konan University, 8-9-1 Okamoto, Kobe, Hyogo 658-8501, Japan \\ 
$^4$ Kavli Institute for the Physics and Mathematics of the Universe (WPI), The University of Tokyo, 5-1-5 Kashiwanoha, Kashiwa, Chiba 277-8583, Japan \\
$^5$ Departamento de Ciencias F\'isicas, Universidad Andres Bello, Avda. Republica 252, Santiago, Chile \\
$^6$ European Southern Observatory, Alonso de C\'ordova 3107, Casilla 19001, Santiago, Chile \\
$^7$ Centro de Modelamiento Matem\'atico, Universidad de Chile, Av. Blanco Encalada 2120 Piso 7, Santiago, Chile \\
$^8$ Instituto de Astrof\'isica La Plata, IALP (CCT La Plata), CONICET-UNLP,
Paseo del Bosque s/n, 1900 La Plata, Argentina \\
$^9$ Facultad de Ciencias Astron\'omicas y Geof\'isicas, Universidad Nacional de La Plata, Paseo del Bosque s/n, 1900 La Plata, Argentina \\
$^{10}$ Las Cumbres Observatory Global Telescope Network, Goleta, CA 93117, USA \\
$^{11}$ Department of Physics, University of California, Santa Barbara, CA 93106-9530, USA \\
$^{12}$ School of Physics and Astronomy, University of Southampton, Southampton, SO17 1BJ, UK \\
$^{13}$ Department of Astronomy and Astrophysics, University of Toronto, 50 St. george Street, Toronto, ON, M5S 3H4, Canada \\
$^{14}$ ITEP, Bolshaya Cheremushkinskaya 25, 117218 Moscow, Russia \\
$^{15}$ Novosibirsk State University, ul. Pirogova 2, Novosibirsk, 630090 Russia \\
$^{16}$ SAI, Moscow University, Universitetski pr. 13, 119992 Moscow, Russia \\
}

\begin{document}

\date{}

\pagerange{\pageref{firstpage}--\pageref{lastpage}} \pubyear{}

\maketitle

\label{firstpage}

\begin{abstract}
We investigate the early-time light-curves of a large sample of 223 type II supernovae (SNe) from the Sloan Digital Sky Survey and the Supernova Legacy Survey. Having a cadence of a few days and sufficient non-detections prior to explosion, we constrain rise-times, i.e. the durations from estimated first to maximum light, as a function of effective wavelength. At restframe $g'$-band ($\lambda_{\mathrm{eff}}=4722$\AA), we find a distribution of fast rise-times with median of $(7.5\pm0.3)$ days. Comparing these durations with analytical shock models of \citet{Rabinak13,Nakar10} and hydrodynamical models of \citet{Tominaga09}, which are mostly sensitive to progenitor radius at these epochs, we find a median characteristic radius of less than 400 solar radii. The inferred radii are on average much smaller than the radii obtained for observed red supergiants (RSG). Investigating the post-maximum slopes as a function of effective wavelength in the light of theoretical models, we find that massive hydrogen envelopes are still needed to explain the plateaus of SNe~II. We therefore argue that the SN~II rise-times we observe are either a) the shock cooling resulting from the core collapse of RSG with small and dense envelopes, or b) the delayed and prolonged shock breakout of the collapse of a RSG with an extended atmosphere or embedded within pre-SN circumstellar material.
\end{abstract}

\begin{keywords}
supernovae -- red supergiants.
\end{keywords}

\section{Introduction}

In the standard picture of single massive star evolution, red supergiant stars (RSG) with zero-age main-sequence mass between approximately 8 and 25 $M_{\odot}$ end their lives with the collapse of their cores (CC) in luminous explosions known as type II supernovae or SNe~II \citep[for a review see e.g.][]{Smartt09rev}. The progenitors retain part of their hydrogen layers prior to explosion, so that their SNe display hydrogen-rich spectra and varying light-curve shape: from linearly declining after maximum, commonly known as SNe~IIL, to plateau supernovae, SNe~IIP, with characteristic post-maximum plateau \citep{Barbon79,Heger03}. The variation of plateau lengths is thought to be mainly produced by the amount and density of hydrogen which produces the recombination wave traveling inwards in mass coordinates \citep[e.g.][]{Litvinova85,Kasen09b,Bersten11}. Recently, the study of large samples of SNe~IIP and SNe~IIL have driven discussion as to whether these two groups actually form a single continuous set \citep[e.~g.][]{Arcavi12,Anderson14,Gutierrez14,Sanders14,Faran14P,Faran14L}. In addition, two other hydrogen-rich SNe are identified: SNe~IIn and SNe~IIb. SNe~IIn are supernovae whose ejecta interact with dense circumstellar material producing distinctive narrow hydrogen emission lines in their spectra and powering the light-curve \citep[e.g.][]{Schlegel90,Taddia13,Moriya11}. SNe~IIb belong to the class of stripped-envelope explosions. They first show hydrogen-rich spectra similar to SNe~IIP but then evolve to hydrogen-deficient SNe~Ib at later times \citep{Filippenko93,Nomoto93,Bufano14}. In this paper we focus on SNe~IIP and SNe~IIL, calling them together ``SNe~II''.

The best evidence for SNe~II originating from RSG stems from their direct identification in pre-explosion images \citep[e.g.][]{Smartt09,Fraser12,Maund13}, although this was predicted by early theoretical works \citep{Grassberg71,Chevalier76,Falk77,Arnett80}. These studies possibly indicate a mass range for SN~II explosions of 8-18$M_{\odot}$, which may suggest that more massive (observed) RSGs do not explode as luminous events. This has been referred to as the ``red supergiant problem'' \citep{Smartt09}. Low upper limits on the initial mass have been inferred from nucleosynthetic yields from nebular spectra analysis \citep{Jerkstrand14,Dessart10a}, however hydrodynamical models yield larger masses \citep{Utrobin08,Utrobin09,Bersten11}. On the other hand, the family of type II SN~1987A-like SNe \citep[e.g.][]{Taddia11,Pastorello12} are well explained with blue supergiants (BSG) progenitors \citep{Shigeyama88,Woosley88,Arnett89}. Compared with radii of RSG \citep[$200\lesssim R/R_{\odot}\lesssim1500$,][]{Levesque05}, BSG are much smaller \citep[$R/R_{\odot}\lesssim200$,][]{Kudritzki08} and are in principle not expected to be end points of single stellar evolution. However the specifics of e.g. convection, mixing, magnetism and binary interaction can allow for such explosions \citep[see][]{Langer12}.  

A key ingredient to study the properties of the supergiants that result in SN~II explosions is the early light-curve prior to the characteristic plateau where hydrogen recombination dominates the SN display. The first electromagnetic radiation emitted by a SN occurs when the shock wave created from CC emerges from the stellar surface \citep[e.~g.][]{Colgate74,Falk77,Klein78,Ensman92}. This shock breakout emission is expected to be of short duration (few hours for typical SNe~IIP) and has been the subject of recent studies as new observing capabilities offer the possibility to detect it. It is brightest in high frequencies, X-ray and ultraviolet (UV), and a few claims of detections exist for some hydrogen-free CC~SNe (SN~2006aj, \citealt{Campana06}; SN~2008D \citealt{Soderberg08}), and for two SNe~IIP with optical data from the Supernova Legacy Survey and near-UV from the Galaxy Evolution Explorer (SNLS-04D2dc, \citealt{Gezari08,Schawinski08}; SNLS-06D1jd \citealt{Gezari08}). The detection of this flash can give tight constraints on progenitor characteristics \citep[e.~g.][]{Tominaga09,Nakar10}. The posterior rise is powered by the shock cooling as the heated expanding stellar envelope diffuses out \citep[e.~g.][]{Grassberg71,Chevalier76}. The study of the early rise of the main light-curve can also give revealing constraints on the progenitor \citep{Gezari10}.

Despite the growing number of nearby SNe~II discovered with high quality data, the early parts of the light-curves are still quite unexplored, yet fundamental to constrain progenitor characteristics. Some constraints on individual well observed nearby SNe~II exist and show that they rise faster than 10 days in $B$-band and more slowly at longer wavelengths \citep{Leonard02b,Leonard02a,Sahu06,Tomasella13}. Recently, \citet{Gall15} analyze the rise-times of 19 low-$z$ SNe~II confirming short rise-times and finding some evidence for a relation between the rise-time and the peak brightness.

In this work we study the early light-curves of a large sample of SNe~II (SNe~IIP and SNe~IIL) from two rolling searches with good cadence and sufficient data prior to explosion: the Sloan Digital Sky Survey-II Supernova Survey \citep[hereafter SDSS-SN,][]{Sako14} and the Supernova Legacy Survey \citep[hereafter SNLS,][]{Astier06,Guy10}. By comparing the data with analytical shock models and hydrodynamical shock models, we are able to constrain specific progenitor properties. A plan of the paper follows: in section~\ref{models} we shortly review the theoretical models used in the analysis, in section~\ref{data} we present the light-curve data, in section~\ref{analysis} the tools we use in the analysis, in~\ref{results} we show our main results and summarize them in~\ref{summary}.

\section{Shock models}\label{models}

\subsection{Analytical shock models}

Analytical models for the shock breakout and immediate evolution of an exploding supergiant star offer the advantage of being more practical than more complicated numerical models for fitting light-curves with varying progenitor parameters in an unexpensive way. However their treatment is simplistic and need to be tested with more sophisticated calculations (see section~\ref{hydro}). We use the analytical models of \citet[][hereafter NS10]{Nakar10} and \citet[][hereafter RW11]{Rabinak11}. These models provide analytic expressions relating the emission and the model parameters such as progenitor mass, radius, and explosion energy. Both models assume pre-explosion density profiles of $\rho(r)\propto(R_* -r)^n$, where $R_*$ is the progenitor radius and the power-law is $n=3$ for radiative and $n=3/2$ for convective envelopes, typical of BSG and RSG stars respectively. The assumed density profiles are only valid for the outermost part of the progenitor structure. The models start post explosion, with that of NS10 including the actual shock breakout, and follow the spherical expansion phase of the stellar envelope until the onset of recombination and/or radioactive decay. They can thus be used in principle, depending on the model parameters, for the first hours up to several days for red supergiants (see eq.~16 in RW11). After this, when recombination sets in, the assumption of constant opacity starts to break down. RW11 include a model extension for varying opacities but we do not use it in this analysis. 

\begin{figure*}%[htbp]
\centering
\includegraphics[width=0.49\linewidth]{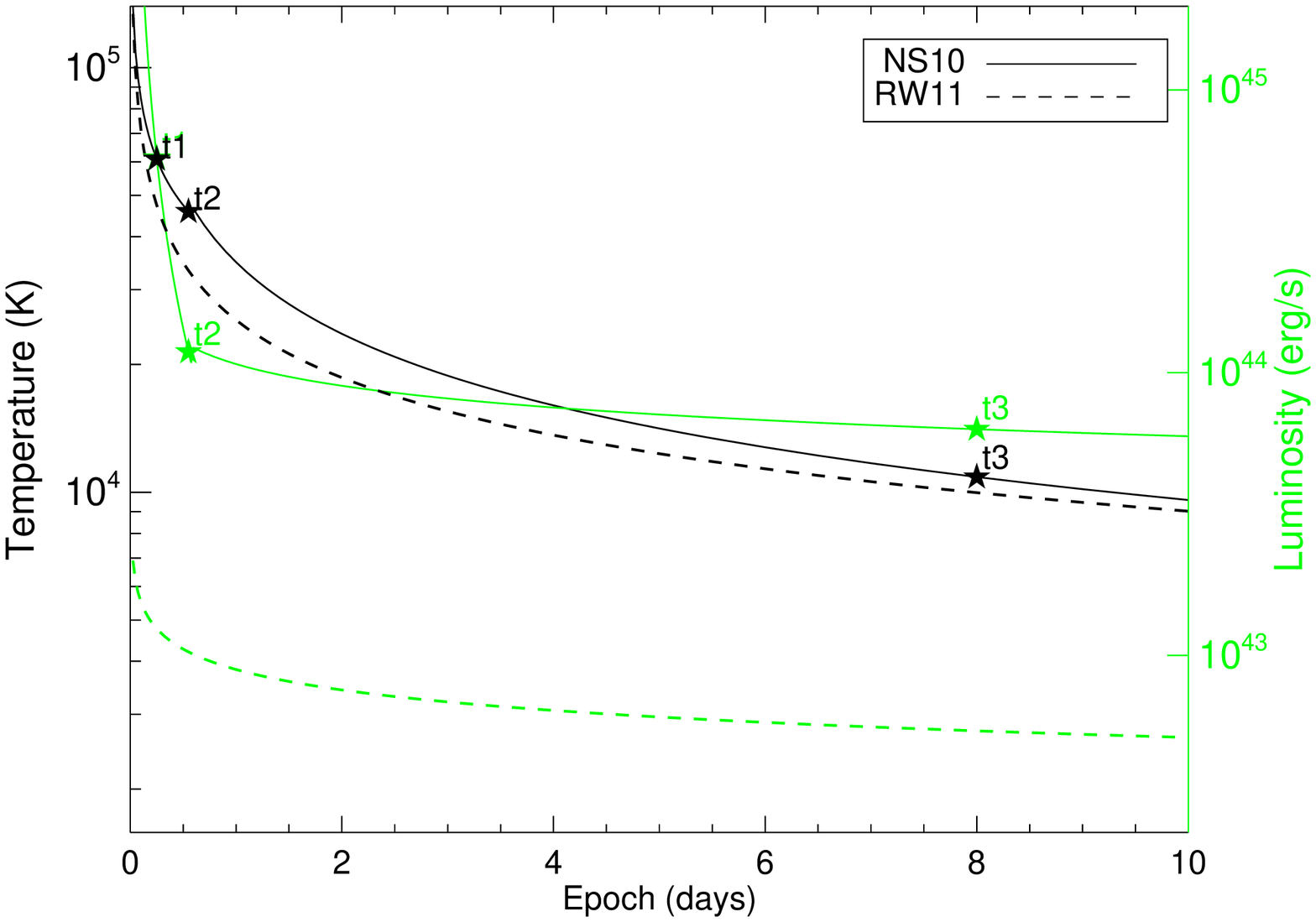}
\includegraphics[width=0.49\linewidth]{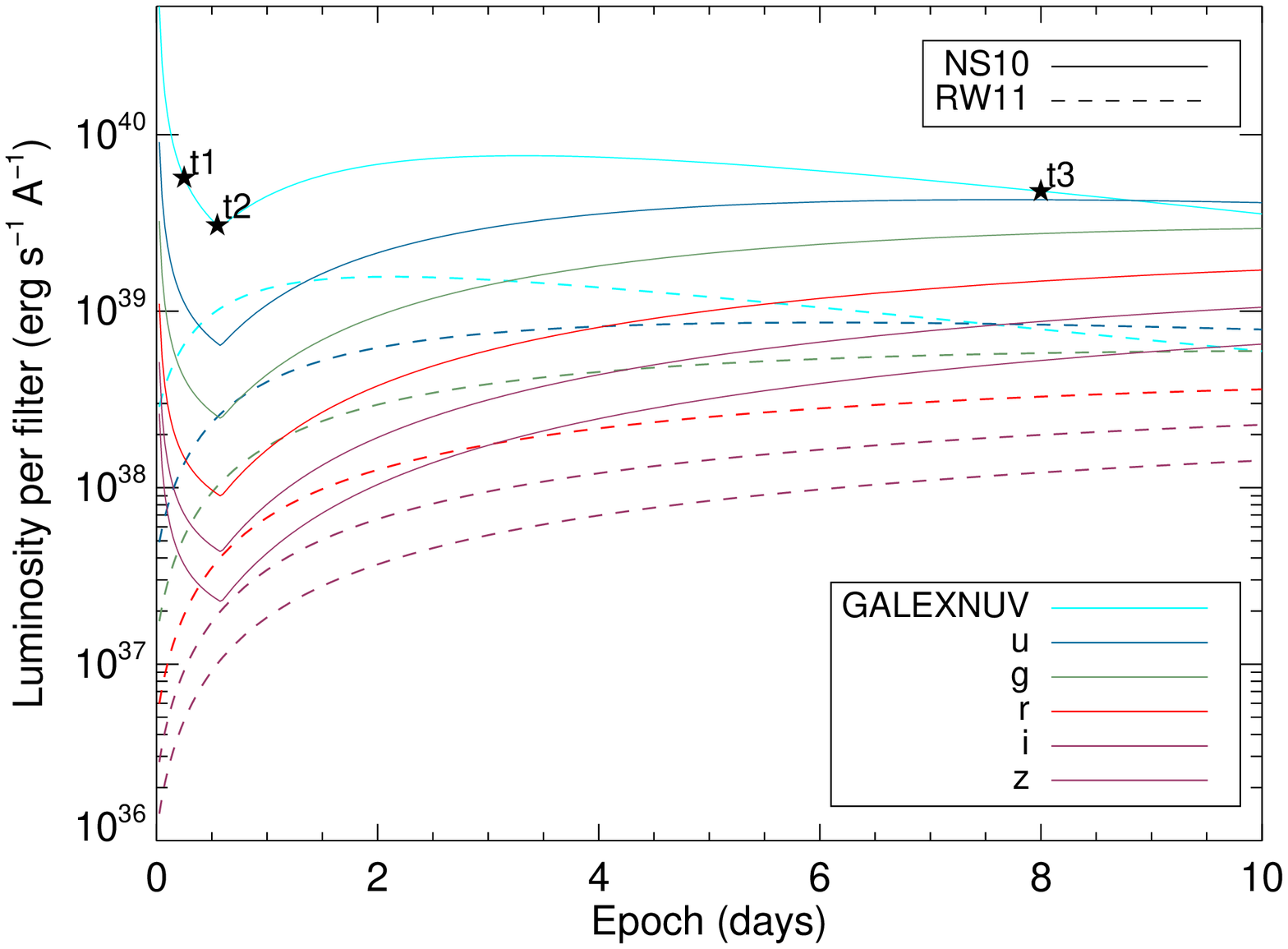}
\caption{\emph{Left}: Early photospheric temperature and bolometric luminosity evolution for models of NS10 and RW11 of a red supergiant progenitor with 500$R_{\odot}$, 15$M_{\odot}$ and 1 foe ($10^{51}$erg). The epoch is with respect to explosion. \emph{Right}: Fluxes in filters NUV$_{\mathrm{GALEX}}$ and Sloan $griz$ for the same models. Stars show three characteristics times (see text).}
\label{modelflux}
\end{figure*}

\begin{figure}%[htbp]
\centering
\includegraphics[width=1\linewidth]{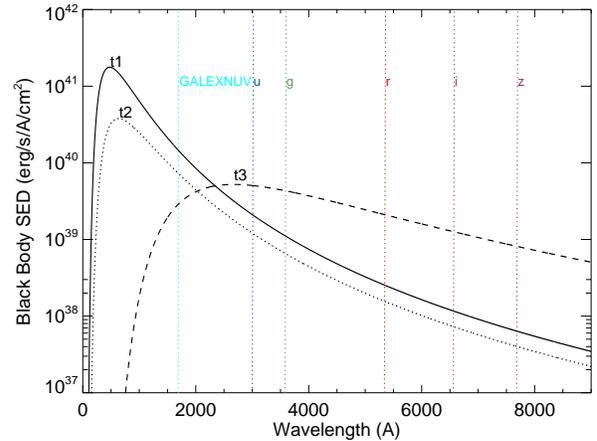}
\caption{Spectral energy distributions at three characteristic times for a standard NS10 model of a red supergiant progenitor with 500$R_{\odot}$, 15$M_{\odot}$ and 1 foe. The mean wavelengths of filters NUV$_{\mathrm{GALEX}}$ and Sloan $griz$ are shown.}
\label{modelsed}
\end{figure}

The analytical models of RW11 and NS10 can be seen in Figure~\ref{modelflux} for a standard scenario with $R=500R_{\odot}$, $M=15M_{\odot}$ and $E=10^{51}$erg (1 foe). The bolometric luminosity and photospheric temperature evolution of the first 10 days are shown, as well as the corresponding prediction for NUV and optical bands. The NS10 models show in all bands a radiation precursor or shock breakout, i.e. an initial peak and decrease at very early phases, followed by the usual rise to maximum light. This behaviour originates from the evolution of the spectral energy distribution as the bolometric luminosity drops together with the temperature as well, so that the tail of the radiation at longer wavelengths is higher at later times due to lower temperatures (see Figure~\ref{modelsed}). The duration and shape of the precursor as well as the rise of the light-curve depend on model parameters, particularly the radius, as will be shown later. RW11 do not model the very early phases of shock breakout.

\subsection{Hydrodynamical models}\label{hydro}

\begin{figure}%[htbp]
\centering
\includegraphics[width=1.0\linewidth]{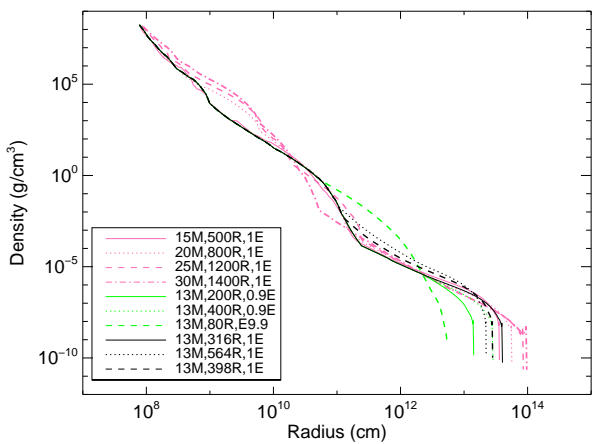}
\caption{Density profiles of standard hydrodynamical models by T09 (pink), models with small radii (green) and dense models (black).}
\label{dens}
\end{figure}

Numerical calculations to provide models for the early evolution of the emission of a SN have been developed by several authors \citep{Falk77,Ensman92,Blinnikov00,Utrobin07,Gezari08,Tominaga09,Bersten11}. They have the advantage with respect to analytical models of treating the pre-SN stellar structure and the physics in a more realistic manner. We focus here on the models by \citet[][herefater T09]{Tominaga09,Tominaga11}. The various progenitor models with varying mass, radii and explosion energy are summarized in Table~\ref{table-tom}, and come from stellar evolutionary models by \citet{Umeda05}. The first set of models are RSG with $R/R_{\odot}=500-1400$ but we also add a second and a third set of models (not present originally in T09) with a modification of the H/He envelope structure. The models in the second set have smaller radii of $R/R_{\odot}=80-400$, and the models in the third set have dense envelopes with intermediate radii of $R/R_{\odot}=300-600$ (see Figure~\ref{dens}). All models assume a metallicity of $Z=0.02$. Higher metallicity leads to more extended pre-SN structures, so its effect on the shock and rise of the light-curve is similar to progenitor radius. Explosive nucleosynthesis follows the method of \citet{Tominaga07} and radiation hydrodynamics is performed with the code STELLA \citep{Blinnikov98,Blinnikov00,Blinnikov06}. This model was successfully used in T09 for SNLS-04D2dc and its GALEX counterpart.

\begin{table}
 \centering
\caption{Progenitor Models from T09: main sequence mass, pre-explosion radius and explosion energy. A fixed metallicty of Z=0.02 is assumed.}\label{table-tom}
 \begin{tabular}{ccc}
\hline
\hline
$M/M_{\odot}$ &  $R/R_{\odot}$ & $E/(10^{51}\mathrm{erg})$ \\
\hline
15 & 500 & 1 \\
20 & 800 & 1 \\
25 & 1200 & 1 \\
25 & 1200 & 9 \\
25 & 1200 & 19 \\
30 & 1400 & 1 \\
\hline
13 & 80 & 0.9 \\
13 & 80 & 9.9 \\
13 & 200 & 0.9 \\
13 & 200 & 9.9 \\
13 & 400 & 0.9 \\
13 & 400 & 9.9 \\
\hline
13 & 316 & 1 \\
13 & 398 & 1 \\
13 & 564 & 1 \\
\hline
\end{tabular}
\end{table}

\section{Data}\label{data}
The data used in this analysis comes from two rolling searches designed primarily to discover type Ia supernovae at intermediate redshifts: the SDSS-SN with SN~Ia redshifts of $0.1<z<0.4$ and the SNLS with SN~Ia redshifts of $0.1<z\lesssim1.1$. These non-targeted searches have the advantage of simultaneously discovering many SNe of different types and sampling light-curves at all epochs, including non-detections prior to explosion and the scarcely covered rise-times of SNe~II. The non-Ia SNe represent an exquisite dataset which to-date has not been fully explored (with a small number of exceptions \citealt{Taddia14,Taylor14,DAndrea10,Bazin09,Nugent06}).

\begin{enumerate}[1.]
\item {\bf SDSS-SN}

The SDSS-SN was one component of the SDSS-II projects using the 2.5m telescope at Apache Point Observatory \citep{Gunn06,York00}. It repeatedly imaged 300 deg$^2$ every 3-4 days in rest-frame in $ugriz$ filters \citep{Fukugita96} with a spectroscopic followup program \citep{Sako08}. In \citet{DAndrea10}, the light-curves of 34 spectroscopically confirmed SNe~IIP in the redshift range $0.027<z<0.144$ were published; and more recently, \citet{Sako14} presented the full dataset of more than 10000 transients discovered during the fall 2005-2007 campaigns. This full dataset comprises 62 spectroscopically confirmed SNe~II (including those from \citealt{DAndrea10}) but it also has many more photometric SN~II candidates with host galaxy spectroscopic \citep{Sako14,Olmstead14} or photometric \citep [DR8,][]{Aihara11} redshifts.

\item {\bf SNLS}

The SNLS \citep{Perrett10} used the deep component of the 3.6m Canada-France-Hawaii Telescope (CFHT) Legacy survey to image four 1 deg$^2$ fields repeatedly every 2-3 days in rest-frame in filters $g_Mr_Mi_Mz_M$, similar to the SDSS passbands \citep{Regnault09}, and with a spectroscopic followup component at various large telescopes \citep{Howell05,Ellis08,Bronder08,Balland09,Walker11}.  A small fraction of the SNe~IIP were used in \citet{Nugent06} for distance estimation analyses. The full dataset of photometric transients consists of more than 6000 objects, for which a fraction also have spectroscopic \citep{Davis03,LeFevre05,Lilly07} or photometric \citep{Ilbert06,Freeman09} redshifts from their host galaxies.
\end{enumerate}

We search for SN~II candidates within the full photometric sample of both surveys. For this, we use a simple identification tool based on the shallowness of post-maximum slopes in different bands \citep[see Gonzalez method in][]{Kessler10} and a posterior visual inspection to eliminate false candidates. With this method we recover most spectroscopically classified SNe~II. However, given that the general spectroscopic classification of the SDSS-SN and SNLS may include other core-collapse SNe such as interacting SNe~IIn and SNe~IIb, we revise all spectroscopic classifications using SNID \citep{Blondin07} and find that a 20-30\% fraction of them (9 for the SDSS-SN and 3 for the SNLS) are not SNe~IIP/L and were not classified so photometrically either. Nonetheless, we caution that there are a small number of events which are photometrically classified as SNe~II, but spectroscopically are consistent with other SN types. This is a source of contamination and is further investigated in section~\ref{results}. 

\begin{figure*}
\centering
\includegraphics[width=0.49\linewidth]{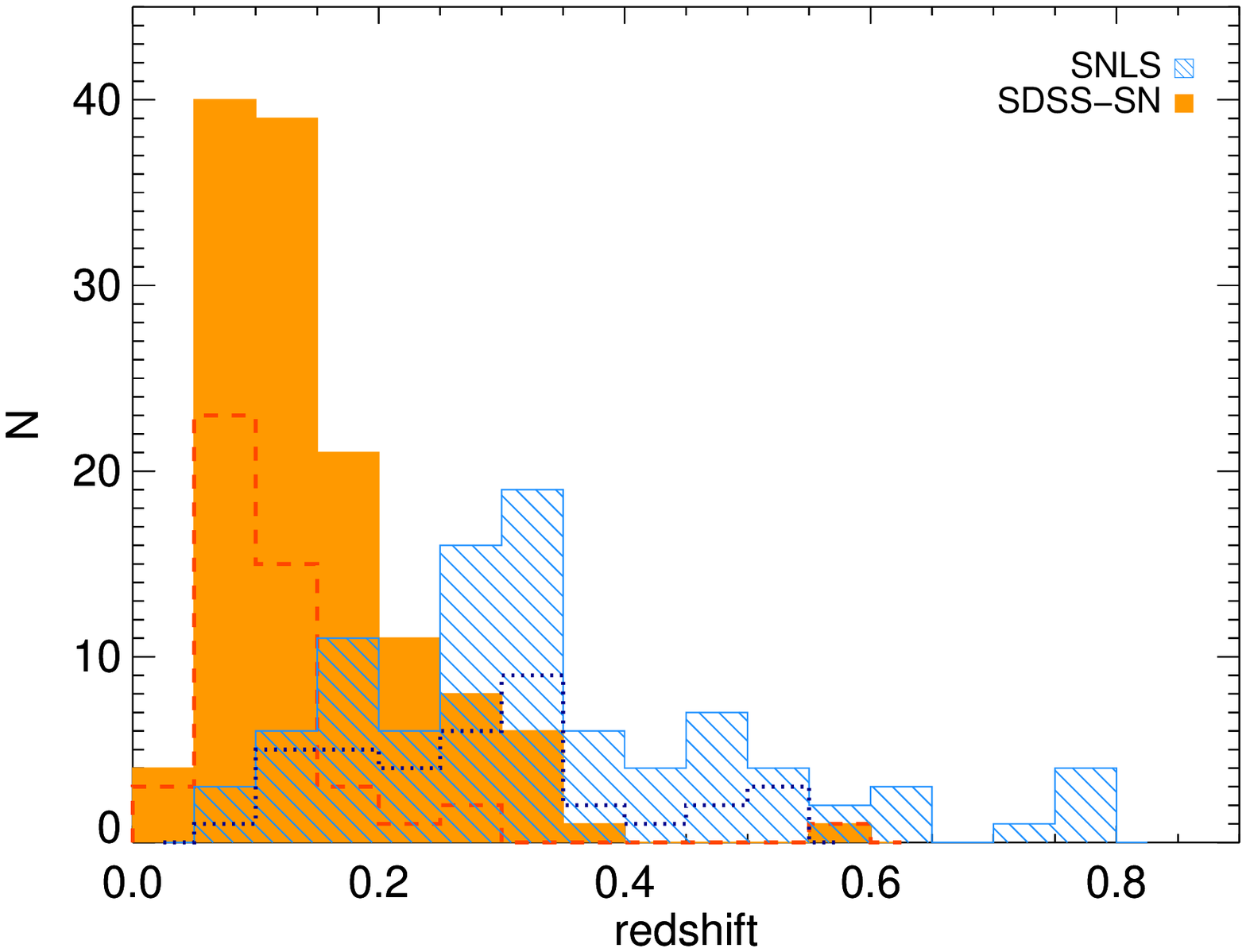}
\includegraphics[width=0.49\linewidth]{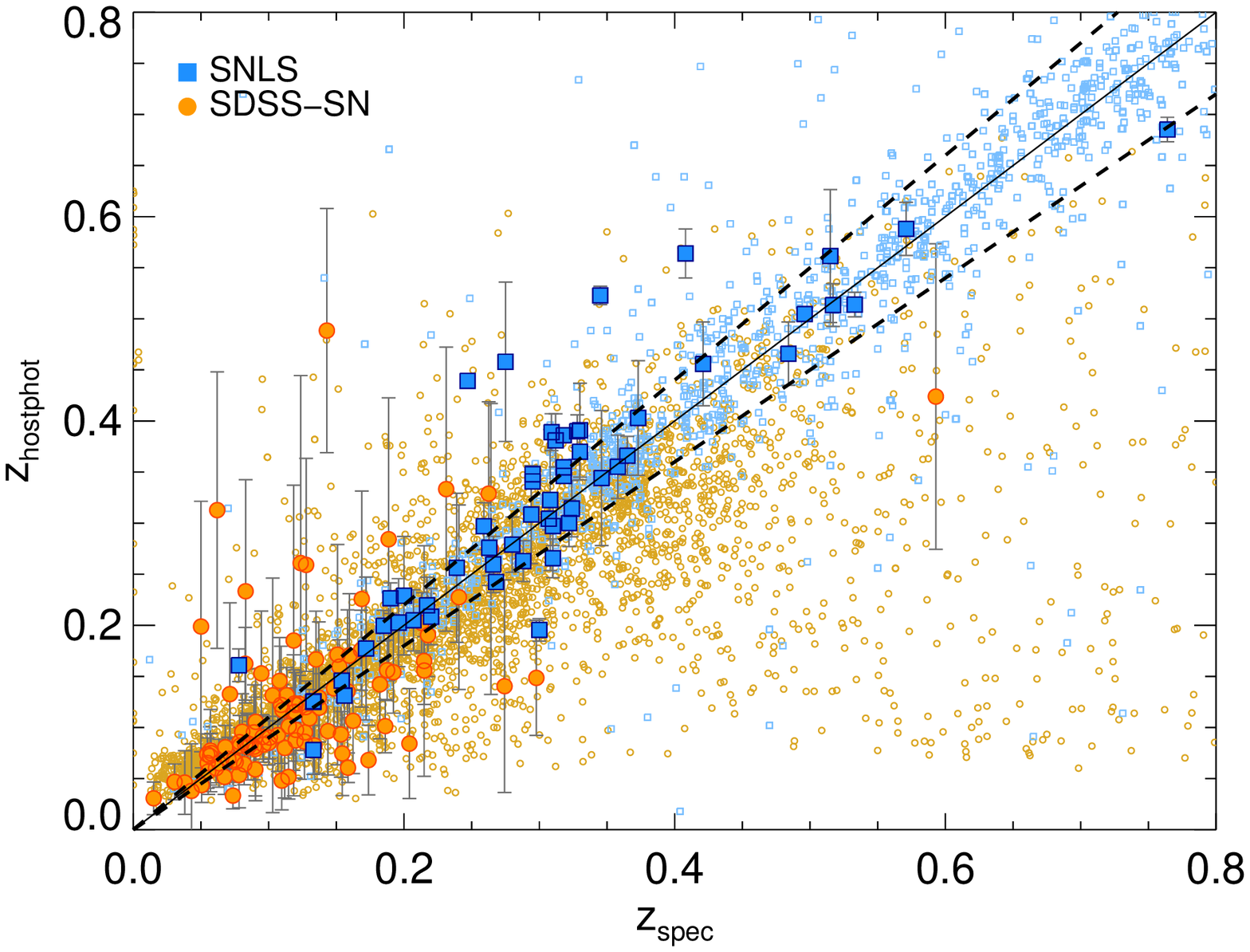}
\caption{Left: Redshift (spectroscopic and host photometric) distribution of silver samples for SDSS-SN (solid orange) and SNLS (diagonal blue) SNe~II. Golden samples are shown in dashed and dotted lines, respectively. Right: Comparison of photometric redshifts from the host with spectroscopic redshifts from the SN or the host for SDSS-SN (orange circles) and SNLS (blue squares) SN~II candidates with spectroscopic information. Smaller open symbols are other transients from each survey. The solid line is a 1:1 relation and dotted lines represent 10\% redshift error.}
\label{zfig}
\end{figure*}

Our final SN~II candidates are divided into a ``golden sample'' which need to fulfill following requirements: a) a spectroscopic redshift from the SN or the host, b) a spectroscopic classification or a good photometric classification as a SN~II and c) at least one data point between the last non-detection and the date of maximum in a given band. Such cuts result in 48 SNe for the SDSS-SN and 38 for the SNLS. A less restrictive ``silver sample'', which includes the golden sample, requires a) a spectroscopic or photometric redshift, b) a spectroscopic classification or a photometric classification and c) at least one non-detection and an estimate of the date of maximum for a given band. As opposed to the golden sample, the silver sample allows photometric redshifts and does not necessarily require data within the rise. The silver sample contains 131 objects for the SDSS-SN in the redshift range $0.1<z<0.6$ and 92 for the SNLS in the redshift range $0.1<z<0.8$. A summary of these samples is presented in Table~\ref{table-sample}. The redshift distributions are shown in Figure~\ref{zfig}. We also show a comparison of SN~II candidates that have both, spectroscopic and photometric, redshifts.

\begin{table}
 \centering
 \caption{Number of SNe in the SDSS-SN and SNLS silver and golden samples. Spectroscopically confirmed SNe~II in parenthesis.}\label{table-sample}
 \begin{tabular}{cccc}
 \hline
 \hline
 \multicolumn{2}{c}{SDSS-SN} & \multicolumn{2}{c}{SNLS} \\
 silver & golden & silver & golden \\
 131(24) & 48(24) & 92(14) & 38(14) \\ 
 \hline
 \end{tabular}
\end{table}

\section{Analysis tools}\label{analysis}
In this section, we present the different tools we use to analyze our photometric data. Instead of performing $K$-corrections \citep{Kim96} to correct light-curve data to the rest-frame, a process which has not been investigated enough for SNe~II, we work directly in the observer-frame taking into account the effective wavelength. This is similar to the approach of \citet{Taddia14} for stripped-envelope SNe.

We work in flux ($f$) space and investigate each filter individually. Some example light-curves are shown in Figure~\ref{lcs}. For each SN and filter, we measure: a) the effective wavelength, b) the rise-time in each filter, and c) the linear slope (in magnitude space) after maximum.

\subsection{Effective wavelengths}

We measure effective wavelengths, \lambdaeff, of each filter for each SN since the objects are found at varying redshift. We use the spectral energy distribution (SED) template at maximum given by \citet{Nugent02}\footnote{\url{https://c3.lbl.gov/nugent/nugent_templates.html}} redshifted to the observer-frame and warped to match the observed colors at maximum \citep[see][]{Hsiao07}. If we use observed spectra at early times instead \citep[e.g. SN2007od 6 days aftex explosion,][]{Gutierrez15}, we obtain consistent effective wavelengths to within 50\AA. Uncertainties in photometric redshifts for some SNe in the silver sample result in errors in \lambdaeff\, of the order of hundreds of \AA. Errors coming from the uncertainties on the observed colors are quite small, on the order of a few \AA. All these errors are added in quadrature and included in the analysis.

\subsection{Rise-times}\label{anal-rise}

\begin{figure*}
\centering
\includegraphics[width=0.49\linewidth]{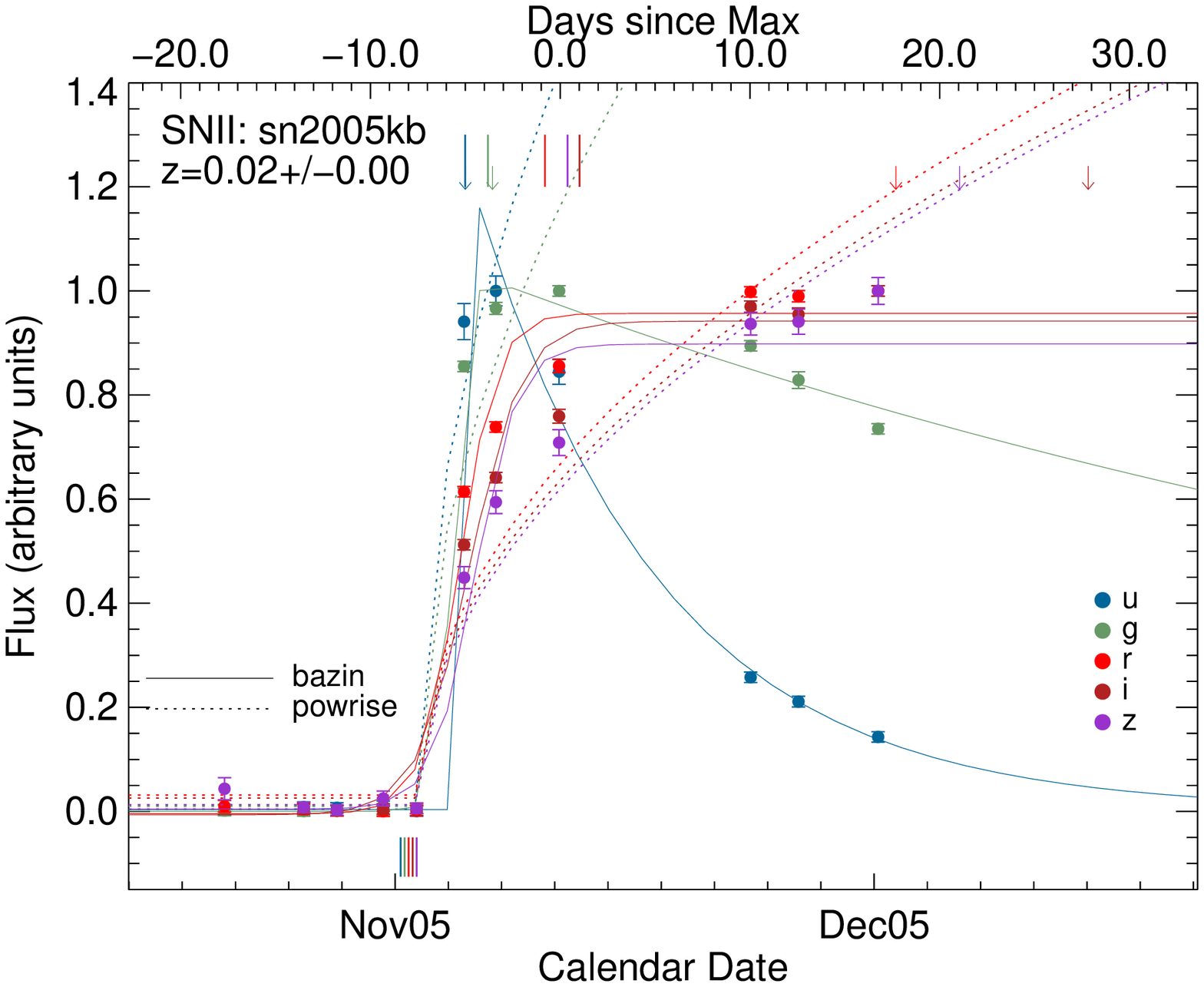}
\includegraphics[width=0.49\linewidth]{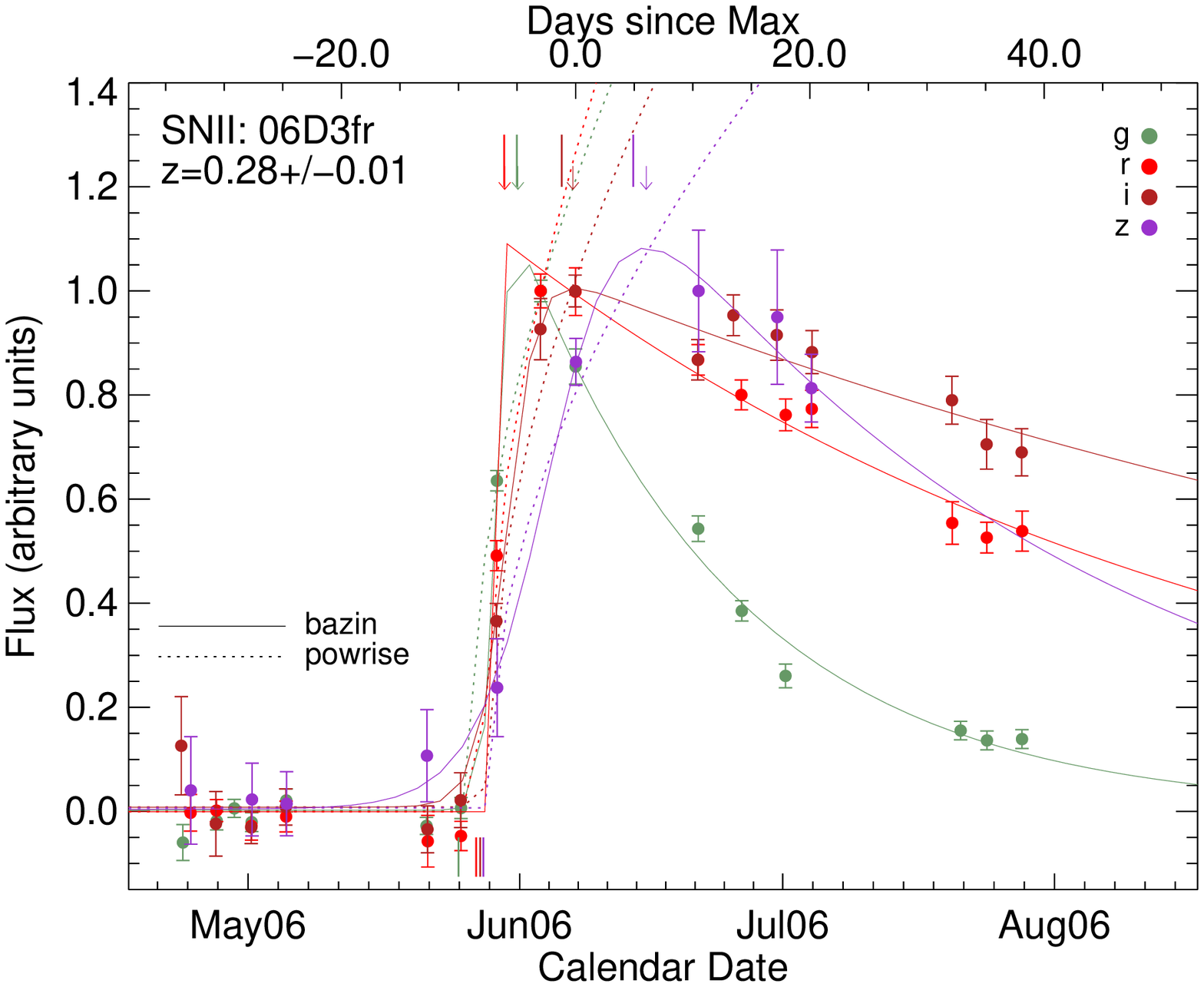}
\caption{Observer-frame light-curve examples of SDSS-SN SN2005kb (left) and SNLS-06D3fr (right). Fluxes normalized to maximum of each filter are shown in different colors. Solid lines are fits to eq.~\ref{eq_baz} \citep{Bazin09} and dotted lines are fits of pre-maximum data to a power law. The small vertical ticks represent the fitted explosion dates, \texp (lower ticks) and the end dates, \tend (upper ticks) for each filter (some have been shifted by half a day for clarity). The arrows indicate the fitted maximum dates. The upper $x$-axis shows the restframe epochs with respect to the maxinmum in the bluest filter ($u$ and $g$ respectively). As an example, we quote the different rise-time measurements in $g$-band (corresponding to $\lambda_{\mathrm{eff}}=4511,3882$\AA) for both SNe, SN~2005kb and SNLS-06D3fr: \fitrise$=4.4\pm0.4,4.4\pm1.2$ and \rise=$4.7\pm0.4,4.5\pm1.2$ days; and in $i$-band (corresponding to \lambdaeff$=7373,6029$\AA): \fitrise=$10.1\pm1.7,6.9\pm1.2$ and \rise=$41.4\pm1.3,7.9\pm1.2$.}
\label{lcs}
\end{figure*}

To measure rise-times, we need to estimate a starting date and an end date of the rise for each band. For the starting date we identify two methods. The first consists of calculating the mid-point between the epoch of last non-detection and the epoch of first detection. Although straightforward, this estimate can lead to systematic effects when obtaining non-detections with different limiting magnitudes. The second method improves the estimate on the explosion date by performing a simple power law fit to the data prior to maximum and including non-detections:
\begin{equation}\label{eq_pow}
f(t)=
\begin{cases}
a(t-t_{\mathrm{exp}}^{\mathrm{pow}})^n & \mathrm{if}\,\, t>t_{\mathrm{exp}}^{\mathrm{pow}} \\
0 & \mathrm{if}\,\, t<t_{\mathrm{exp}}^{\mathrm{pow}} \\
\end{cases}
\end{equation}
with the error obtained directly from the fit. The explosion date obtained in this way, \texp, does not change significantly with the range of epochs used in the fit. The power index $n$, on the other hand, is affected by this limit and a lower upper range is preferred to properly estimate the early shape of the light-curve. To estimate the power-law, we therefore use data up to half of the flux at maximum. 

For the end date of the rise, we can use the date of maximum light in each band. Instead of directly taking the observed maximum which depends strongly on the cadence of the observed SN and generally leads to over-estimations in the case of short rise-times, we do a light-curve fit to the phenomenological function introduced by \citet{Bazin09}:

\begin{equation}\label{eq_baz}
f(t)=A\frac{e^{-(t-t_0)/\tau_{fall}}}{1+e^{(t-t_0)/\tau_{rise}}}+B,
\end{equation}
where $A$, $B$, $t_0$, $\tau_{fall}$ and $\tau_{rise}$ are free parameters. We can get a date of maximum, \tmax,  from the derivative of this function: 
\begin{equation}\label{eq_bazmax}
t_{\mathrm{max}}^{\mathrm{Baz}}=t_0+\tau_{rise}\log\left(\frac{-\tau_{rise}}{\tau_{rise}+\tau_{fall}}\right),
\end{equation}
and an error propagated from the $1\sigma$ uncertainties on the parameters from the fit. 

In some cases, the date of maximum flux in the redder bands can occur during the plateau, much later than the end of the rise and much later than the maximum in bluer bands (see Figure~\ref{lcs}). Although the time between explosion and maximum light is easily estimated, and has been used in the past to measure rise-times of nearby SNe~II, it can be strongly affected by other physical processes like hydrogen recombination at longer wavelengths. To avoid over-estimating rise-times, we use the derivative of the same phenomenological function and find the epoch at which it becomes lower than 0.01mag/day, \tend, in a similar fashion to \citet{Gall15}. This end date is similar to the date of maximum at shorter wavelengths but avoids possible rising plateaus in redder bands. The error is calculated by propagating the errors on the fit parameters. Some examples of these estimates and fits are shown in Figure~\ref{lcs}. We thus present rise-time measurements based on the fitted explosion date, \texp, and the end date from the Bazin function, \tend:
\begin{equation}
 t_{\mathrm{rise}}^{\mathrm{end}}=t_{\mathrm{end}}^{\mathrm{Baz}}-t_{\mathrm{exp}}^{\mathrm{pow}}.
\end{equation}

For comparison with previous works, we also show the rise-times based on the maximum date: 
\begin{equation}
t_{\mathrm{rise}}^{\mathrm{max}}=t_{\mathrm{max}}^{\mathrm{Baz}}-t_{\mathrm{exp}}^{\mathrm{pow}}. 
\end{equation}

As will be shown, our main conclusions are not affected by the use of different measurements of the starting date end dates for the rise-time calculation.

\subsection{Post-maximum slopes}
We measure a post-maximum slope in each filter by doing a linear fit in magnitudes to the data after maximum up to 80 days past maximum. We require at least three epochs to ensure a proper fit. This gives a general quantitative idea of the shape of the light-curve after maximum. It is reminiscent of the ``s2'' plateau slope of \citet{Anderson14} except we do not fit a primary ``s1'' decline.

\section{Results}\label{results}

\subsection{Rise-times versus effective wavelength}

\begin{figure*}%[htbp]
\centering
\includegraphics[width=0.49\linewidth]{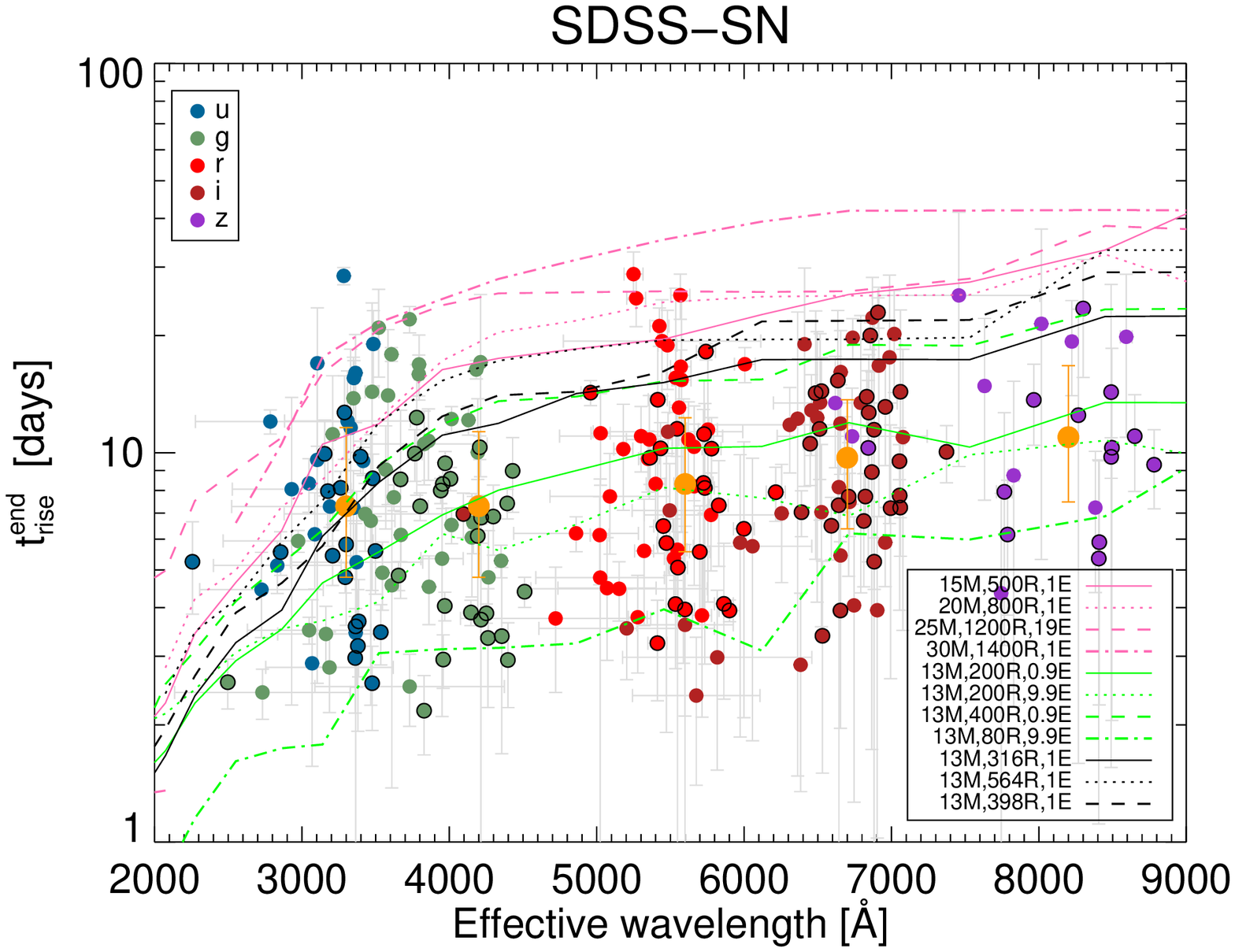}
\includegraphics[width=0.49\linewidth]{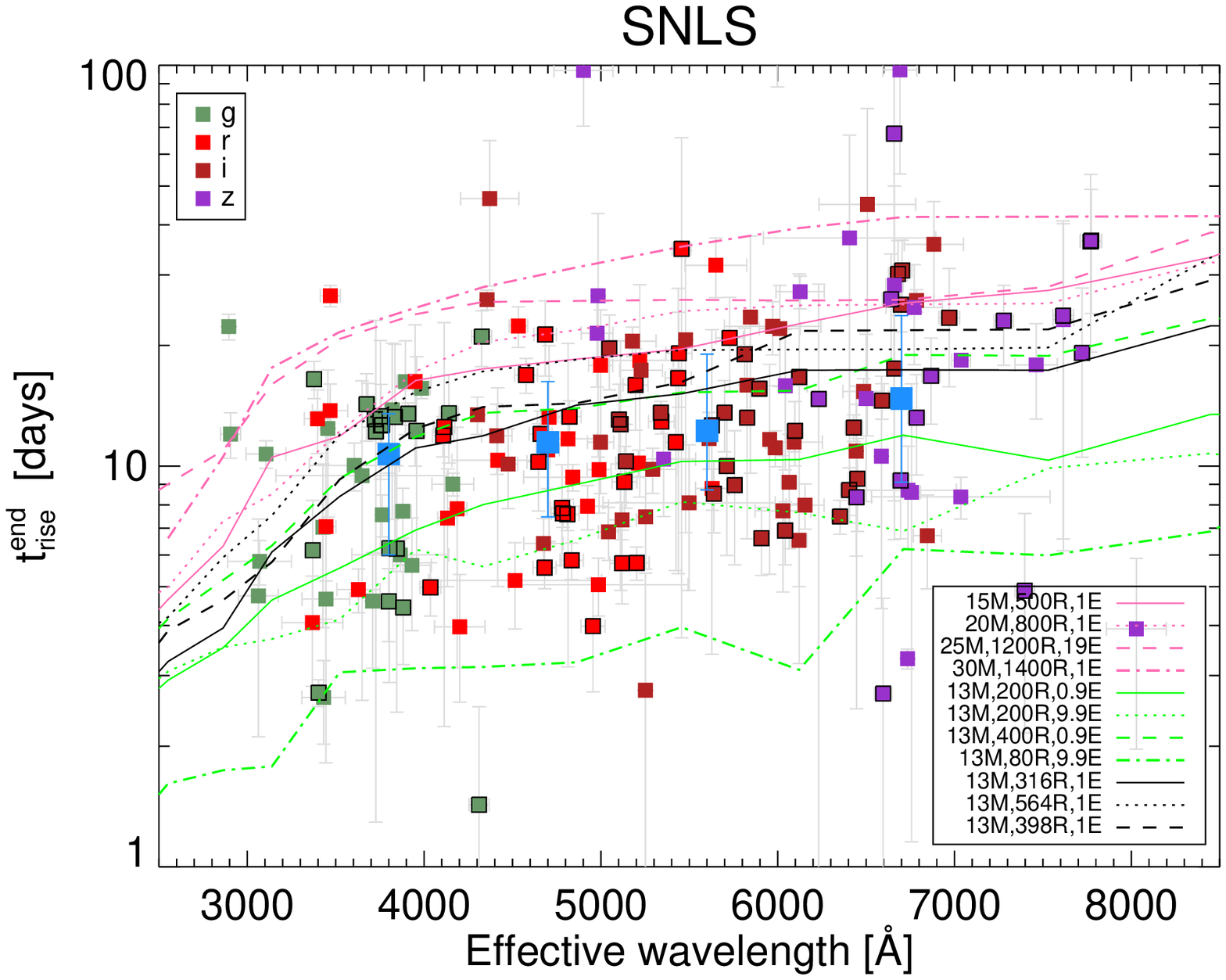}
\caption{Calculated rise-times, $t_{\mathrm{rise}}^{\mathrm{end}}$, for the SDSS-SN (left) and SNLS (right) samples as a function of effective wavelength. Each point is the rise-time of a SN at a given band. All SNe from the silver sample are shown and the SNe from the golden sample have black contours. Different colors indicate the observed filter and larger symbols denote the median values for characteristic wavelengths. Some representative hydrodynamical models by T09 are shown: standard models of large radii in pink, models of small radii in green and dense models in black.}
\label{fitrise-effw}
\end{figure*}

\begin{figure*}%[htbp]
\centering
\includegraphics[width=0.49\linewidth]{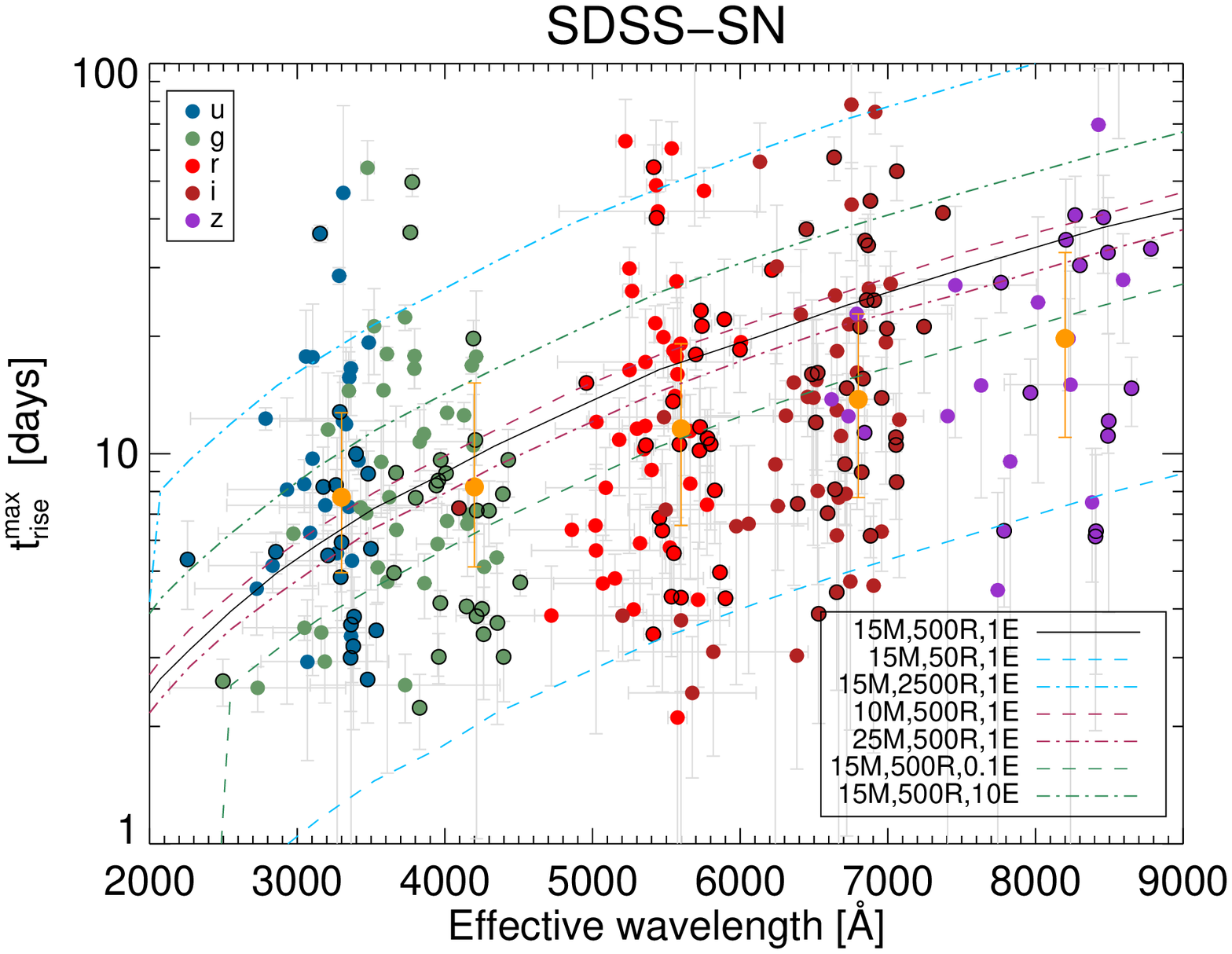}
\includegraphics[width=0.49\linewidth]{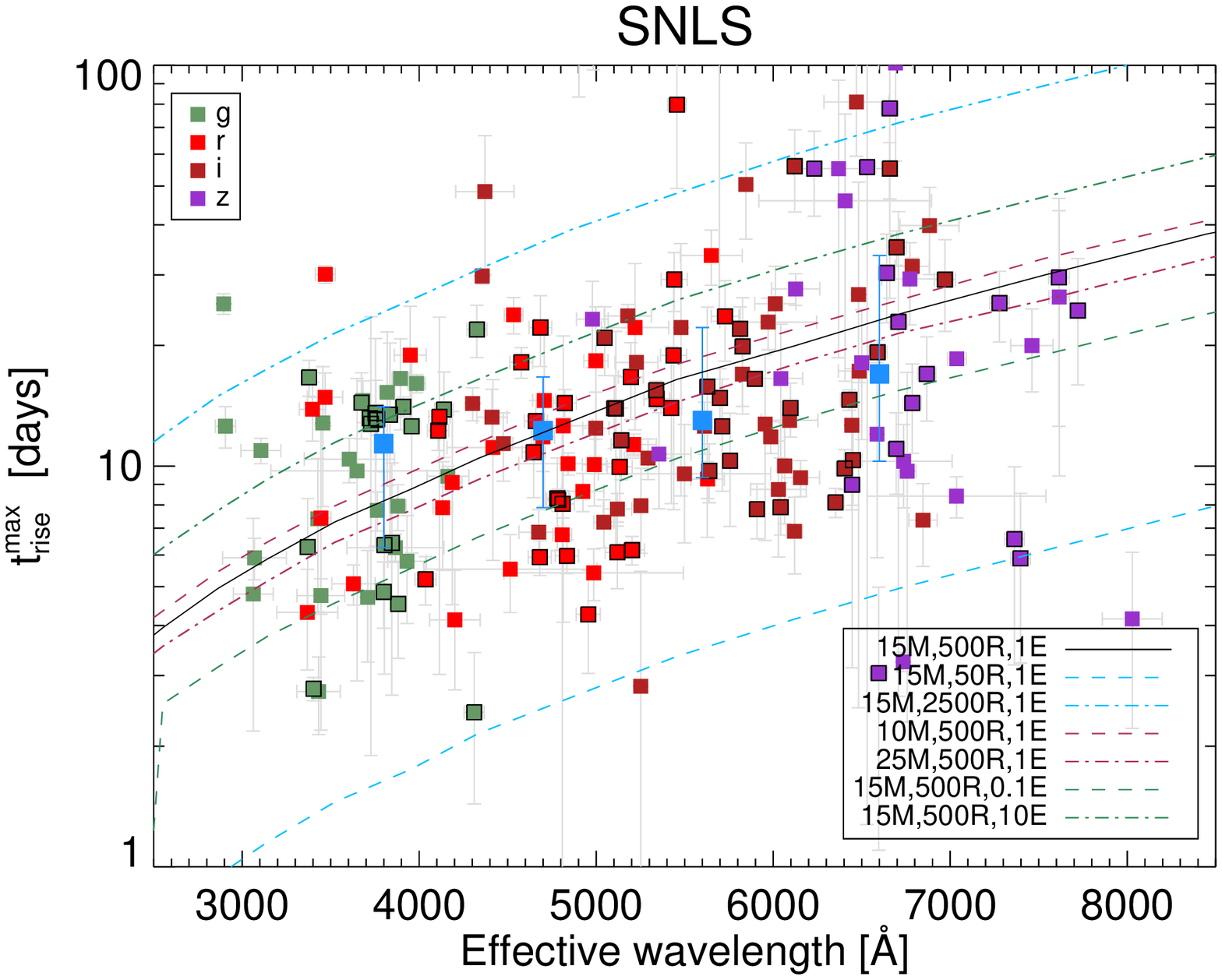}
\caption{Calculated maximum rise-times, \rise, for the samples of SDSS-SN (left) and SNLS (right) as a function of effective wavelength. Each point is the rise-time of a SN at a given photometric band. Black contours indicate SNe from the golden samples, the rest are from the silver samples. Different colors indicate the observed filter and larger symbols denote the median values for characteristic wavelengths. Some representative analytical models by NS10 are shown for comparison: a RSG progenitor with $15M_{\odot}$, $500R_{\odot}$ and 1foe explosion energy (solid black line), varying the mass to 10 or 25$M_{\odot}$ (dashed and dot-dashed maroon), or varying the radius to 50 or 2500$R_{\odot}$ (dashed and dot-dashed blue), or varying the energy to 0.1 or 10 foe (dashed and dot-dashed green).}
\label{exprise-effw}
\end{figure*}

We present the rise-times as a function of effective wavelength for the SDSS-SN and SNLS, and compare them to analytical and hydrodynamical models. For consistency, we calculate the rise-times for data and models in the same manner. Figures ~\ref{fitrise-effw} and ~\ref{exprise-effw} show the rise-times calculated with \tend\, and \tmax, respectively, as a function of effective wavelength. We show in colors the observer-frame filters from which we measured each rise-time, so that typically a SN with good coverage has 4 (SNLS) or 5 (SDSS-SN) values in these plots, each at different wavelength. The observed color of the SN influences the effective wavelength but it is mostly the redshift that gives rise to the diversity for a given photometric band. Since the SNLS spans a larger redshift range, the dispersion on \lambdaeff\, for a given filter is larger and one can see overlap between two filters for the same wavelength. A summary of the rise-times is presented in Table~\ref{table-rise}.

\begin{table*}
 \centering
\caption{Median and deviation on the median for two rise-times measurements, \fitrise\, and \rise, for the SDSS-SN and SNLS silver and golden samples at different effective wavelengths.}\label{table-rise}
 \begin{tabular}{|c|cc|cc|cc|cc|}%c|c|c|c|}

Effective wavelength &  \multicolumn{4}{c|}{SDSS-SN} &  \multicolumn{4}{c|}{SNLS} \\
(\AA) & \multicolumn{2}{c}{silver} & \multicolumn{2}{c|}{golden} & \multicolumn{2}{c}{silver} & \multicolumn{2}{c|}{golden} \\
 &\fitrise(d) &\rise(d) &\fitrise(d) &\rise(d) &\fitrise(d) &\rise(d) &\fitrise(d) &\rise(d) \\
\hline
3300 & $7.3^{+4.3}_{-2.5}$ & $7.7^{+5.0}_{-2.8}$  & $5.8^{+2.8}_{-2.1}$ & $5.9^{+3.0}_{-2.1}$ & -- & -- & -- & -- \\
3800 & -- & -- & -- & -- &  $10.7^{+2.8}_{-4.7}$ &  $11.4^{+2.7}_{-5.1}$ & $12.2^{+1.3}_{-6.1}$ & $13.0^{+0.9}_{-6.7}$ \\
4200 & $7.3^{+4.0}_{-2.5}$  & $8.2^{+7.0}_{-3.1}$  & $6.1^{+2.5}_{-2.3}$ & $6.9^{+2.8}_{-2.9}$ & -- & -- & -- & -- \\ 
4700 & -- & -- & -- & -- &  $11.5^{+4.8}_{-4.0}$  &  $12.3^{+4.4}_{-4.4}$ & $12.0^{+1.6}_{-5.8}$ & $13.0^{+2.5}_{-6.5}$ \\
5600 & $8.3^{+4.0}_{-2.8}$  & $11.5^{+7.0}_{-4.9}$ & $7.9^{+3.6}_{-2.9}$ & $10.6^{+7.9}_{-4.2}$ &  $12.3^{+6.8}_{-3.5}$ &  $14.0^{+9.8}_{-4.2}$  & $12.6^{+4.1}_{-4.1}$ & $14.0^{+8.1}_{-4.2}$ \\
6700 & $9.7^{+4.0}_{-3.3}$ & $13.6^{+9.2}_{-5.9}$ & $8.1^{+3.4}_{-1.6}$ & $12.0^{+11.2}_{-4.0}$ &  $14.7^{+9.0}_{-5.6}$  &  $18.5^{+16.7}_{-8.2}$ & $14.6^{+8.9}_{-5.5}$ & $19.2^{+15.9}_{-8.9}$ \\
8200 & $11.0^{+5.8}_{-3.5}$ & $19.3^{+13.5}_{-8.3}$ & $10.1^{+3.6}_{-2.3}$ & $20.9^{+13.3}_{-9.9}$ & -- & -- & -- & -- \\
\end{tabular}
\end{table*}

The bulk of the rise-times, \fitrise, lies below 10 days for wavelengths less than 4000\AA\, and increases to about 12-15 days beyond 7000\AA. If one uses the maximum rise-times, \rise, they are of the order of 15-20 days beyond 7000\AA. The population of SNe~II clearly has steep rise-times, which are significantly faster than those estimated for stripped-envelope SNe (12-23 days in $r/R$ \citealt{Taddia14}) or standard thermonuclear SNe~Ia (16-25 days in $B$, \citealt{Firth14,Gonzalez12,Hayden10}). This is a possible indication of pure shock cooling of an extended progenitor star powering the early light-curve of SNe~II, as opposed to compact stripped-envelope SNe or SNe~Ia which are generally caught when the light-curve is already powered by radioactivity, e.g. the shock cooling for Wolf-Rayet stars is around 100 seconds \citep{Kistler13} as opposed to the characteristic decay time of 8.8 days for $^{56}$Ni. Previous works of nearby SNe have already shown that individual SNe~II have sharp rises to maximum: SN~1999em rises in 6, 8 and 10 days in $UBV$\citep{Leonard02a}, SN~1999gi rises in 8 and 12 days in $BV$\citep{Leonard02b}, SN~2004et rises in 9, 10, 16, 21 and 25 days in $UBVRI$ \citep{Sahu06}, SN~2012aw rises in less than 3 days in UV and in 8, 11, 15, 22 and 24 days in $UBVRI$ and SN2012A rises in 10, 12, 17 and 24 days in $BVRI$ \citep{Tomasella13}. All these values are consistent with our measurements.

In the figures we also show several analytical models from NS10 and hydrodynamical models from T09. We see that for both, SDSS-SN and SNLS, the standard models of T09 from RSG progenitors with radii 500-1400\rsun\, predict long rise-times and therefore lie consistently above the measured rise-times (pink lines in Figure~\ref{fitrise-effw}). In the case of a second set of new hydrodynamical models with small radii, 80-400 \rsun\, (green lines), we see that the predicted rise-times are in better agreement with the observed rise-times. This is clear evidence for smaller radii than typical assumed RSG radii.

With the analytical models, we see a similar prediction: the rise-times of a 500 \rsun\, progenitor are roughly consistent with the SDSS-SN and SNLS maximum rise-times for NS10 (see Figure~\ref{exprise-effw}). Furthermore, we show other models varying the mass, radius and explosion energy and find that the largest variation in rise-time duration can only be accounted for by radius. As a matter of fact, the luminosity and temperature in their model equations scale more strongly with radius than with other parameters.

\subsection{Distribution of rise-times}\label{dist}
\begin{figure}%[htbp]
\centering
\includegraphics[width=1.\linewidth]{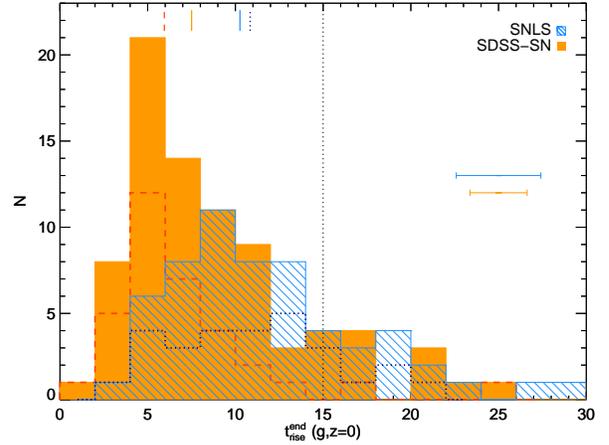}
\caption{Histogram of \fitrise\, rise-times at rest-frame $g$-band (\lambdaeff=4722\AA) for the silver samples of the SDSS-SN (solid orange) and SNLS (diagonal blue). The golden samples are shown in dashed and dotted lines, respectively. Median MC values are indicated with ticks at the top: $7.5^{+7.1}_{-3.6}$ and $6.0^{+3.8}_{-2.8}$ days for the silver and golden samples of the SDSS, and $10.3^{+7.8}_{-4.4}$ and $10.8^{+5.8}_{-5.1}$ days for the silver and golden samples of the SNLS, where the quoted uncertainties contain 84\% of the distribution. The vertical dotted line indicates the limit of long rise-time SNe (see section~\ref{longrise}). The horizontal error bar represent the typical median error on the rise of 2.5 (SNLS) and 1.6 (SDSS-SN) days.}
\label{interpexprise}
\end{figure}

We find that there is a distribution of rise-times for a given wavelength that is larger than the measured rise-time uncertainties. At \lambdaeff$<$4000\AA, the deviation on the median (MAD) is about 4 days. This indicates that besides their brightness and post-maximum decline \citep{Anderson14}, SNe~II constitute a heterogeneous sample during the rising part of their light-curve as well. To further study this, we interpolate the rise-times (\fitrise) of each SN to the common effective wavelength of the SDSS $g$-band at rest-frame, i.e. 4722\AA, and plot the distribution in Figure~\ref{interpexprise}. One can see that the majority of SNe~II have rise-times below 10 days at this wavelength although a population of objects with long rise-times exists and is investigated in next section. To test how our large uncertainties in the rise-times affect this distribution, we perform a Monte Carlo analysis of 100 realizations in which each rise-time is randomly shifted according to its one $\sigma$ uncertainty. We then calculate the mean of all simulations to obtain a characteristic rise-time of $7.5\pm0.4$ ($6.0\pm0.3$) and $10.3\pm0.4$ ($10.7\pm0.7$) days for the silver (golden) samples of the SDSS-SN and SNLS, respectively. If we take both surveys together we obtain $8.9\pm0.2$ and $7.5\pm0.3$ days for silver and golden samples. We also calculate for each simulation the lower and upper rise-times for which 84\% of the population is contained and obtain a final mean range of 4.5-16.6 days. As mentioned in the previous section, all nearby SNe have rise-times of 8-11 days in $B$-band (\lambdaeff$=4391$\AA), in good agreement with this distribution. 

\begin{figure*}%[htbp]
\centering
\includegraphics[width=0.49\linewidth]{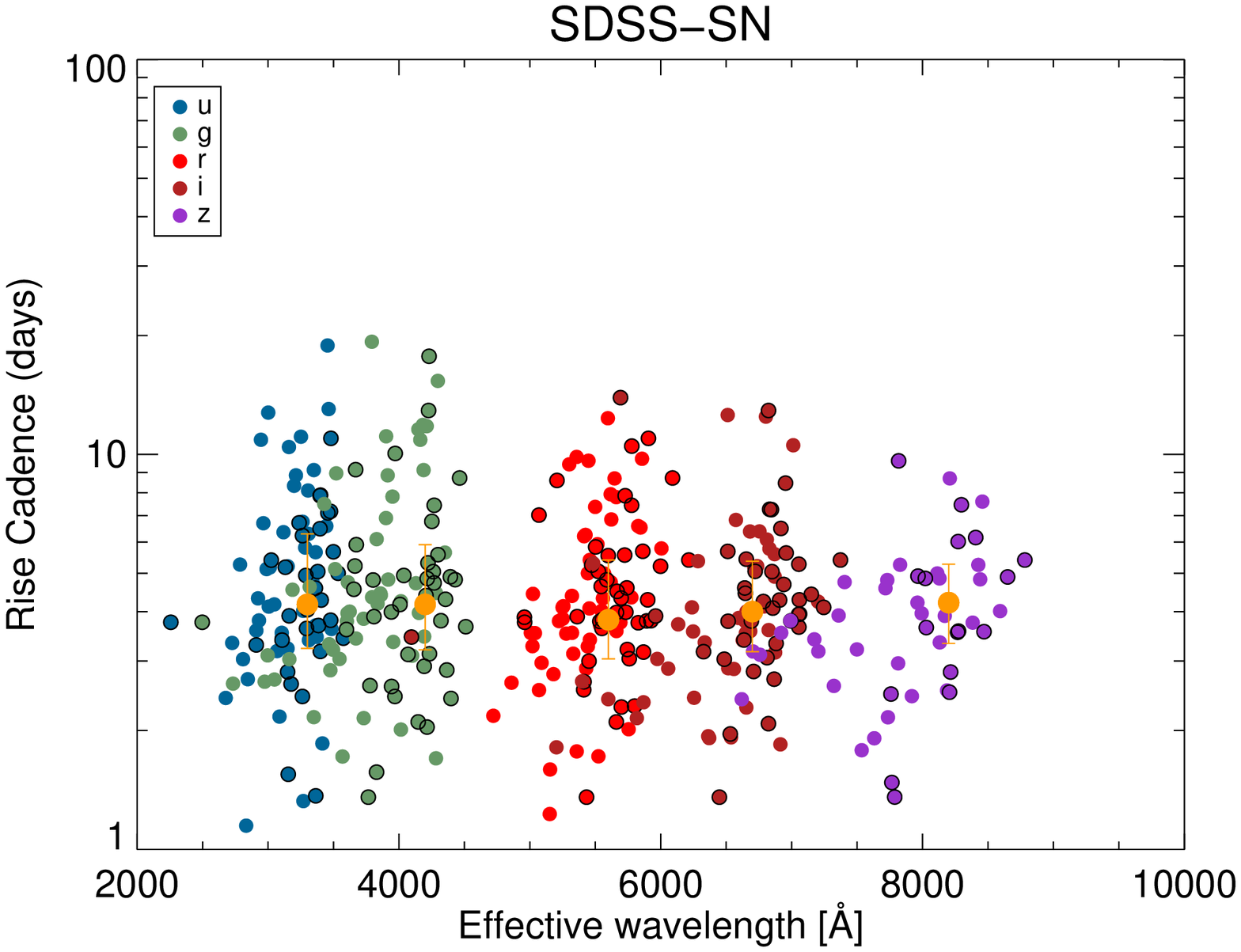}
\includegraphics[width=0.49\linewidth]{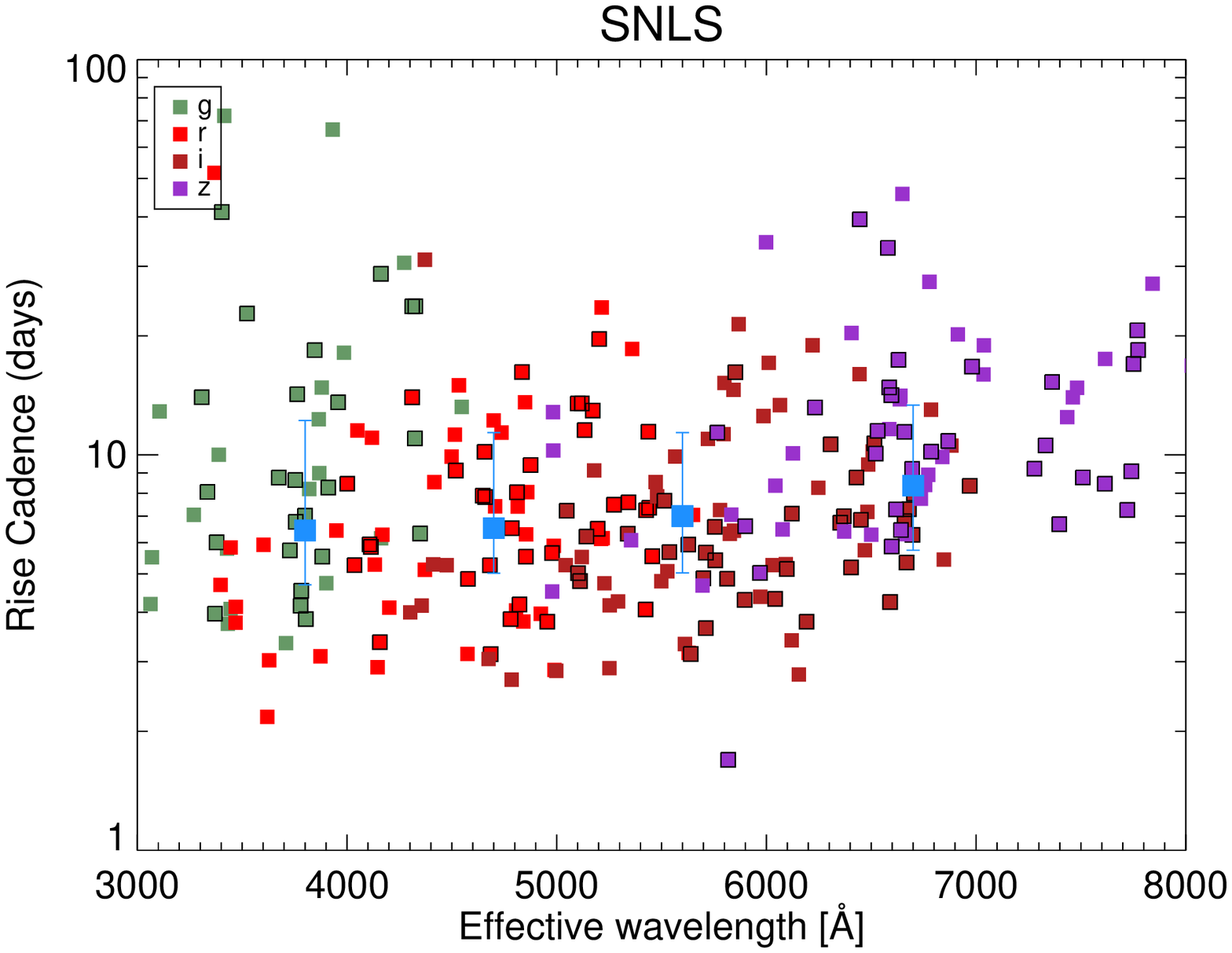}
\caption{Cadence as a function of effective wavelength for the SDSS-SN (left) and SNLS (right). Each point is the rise cadence a SN at a given filter. SNe from the golden sample have black contours. Different colors indicate the observed filter and larger symbols denote the median values for characteristic wavelengths.}
\label{cadence-effw}
\end{figure*}

The large discrepancy between both surveys is found independently of the way we calculate the rise-time. It is due to the difference in cadence for both surveys. In figure~\ref{cadence-effw} we show the measured cadence during the rise for each SN light-curve in our samples. The cadence is the mean rest-frame duration between data points calculated during the rise for a given SN light-curve. We obtain very different average cadences of 4.1 and 8.2 days for the SDSS-SN and SNLS. The mean cadence of the SNLS is of the order of the rise-time measure itself. Since we require at least one point during the rise to calculate the time of explosion in the power-law eq.~\ref{eq_pow}, there will be a lack of objects with short rise-times, as is clearly seen in Figure~\ref{interpexprise}. If we use the midpoint between last non-detection and the first detection to estimate the explosion date instead, we are able to use more SNe, but the rise-times for the SNLS are still consistenly larger due to the long time range without any detection (see simulation in appendix~\ref{syst}). Thus, we believe that the rise-time distribution of the SDSS-SN is more complete and we stress the possibility that there might be even shorter rise-times that are not probed with the cadence of the SDSS-SN. The main conclusions of possible small radiie are not affected by this systematic; in fact they are strengthened.

\subsection{Long rise-time SNe}\label{longrise}

\begin{figure*}%[htbp]
\centering
\includegraphics[width=0.49\linewidth]{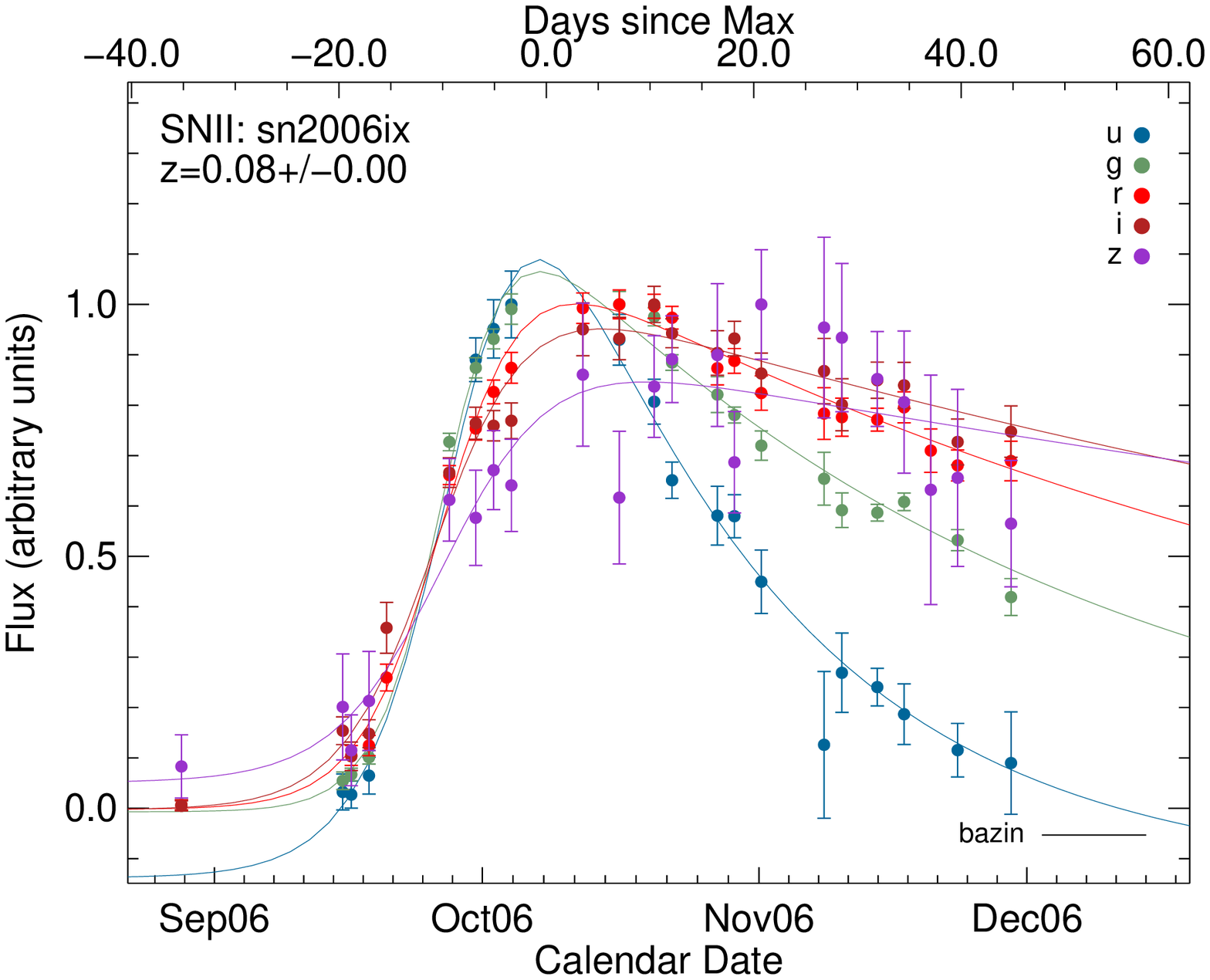}
\includegraphics[width=0.49\linewidth]{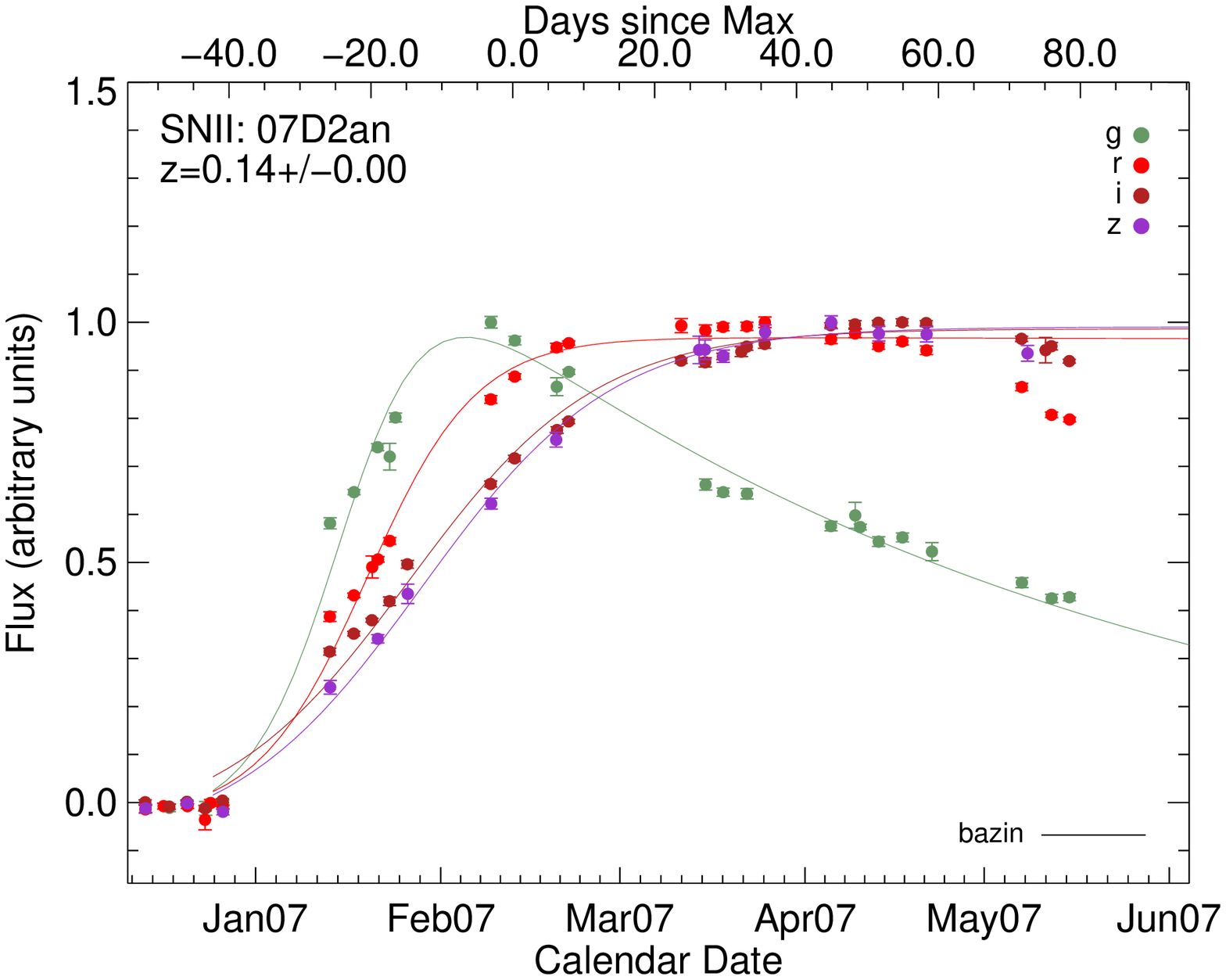}
\caption{Observer-frame light-curve examples of two long rise-time SNe: SDSS-SN SN2006ix, possibly a SN~IIn (left), and SNLS 07D2an, a SN~1987A-like SN~II (right). Fluxes normalized to maximum of each filter are shown in different colors. Solid lines are fits to equation 1  of \citet{Bazin09}. The upper $x$-axis shows rest-frame epochs with respect to maximum in the bluest band ($u$ and $g$ respectively).}
\label{LC-IIn}
\end{figure*}

Despite the fast rise-times found for most of our SNe, we find a fraction of SNe that have much longer rise-times, i.e. larger than 15 days (see Figure~\ref{LC-IIn}). Such different rise durations could indicate that these SNe are not really normal SNe~II but are powered by different energy sources like radioactivity for peculiar SNe~II such as SN~1987A \citep{Utrobin11,Pastorello12} or interaction for SNe~IIn, or even more exotic events such as the super-luminous SNLS-07D2bv \citep{Howell13}. Some of these different types may have shallow slopes like normal SNe~II and could be selected with our classification technique that is based only on post-maximum photometry. To investigate this further, we take a closer look at the SNe originally classified as SNe~II based on post-maximum slopes that have $g$-band rest-frame rise-times at least one sigma above 20 days. We indeed find that one of these was rejected based on its spectrum being consistent with SN~IIn showing narrow absorption H$\alpha$ lines (SN2006ix). Even though it was classified as SN~II photometrically, its long rise-time actually reveals that it is not a standard SN~II. Of all spectroscopically confirmed long rise-time SNe, the only one that is clearly a SN~II is SNLS-07D2an, but the spectrum, particularly the H$\alpha$ P-Cygni profile, is much more consistent with that of peculiar SN~II SN~1987A. We note that this SN has much shallower post-maximum slopes, characteristic of SNe~IIP, as opposed to other SN~1987A-like which decay faster. 

This provides strong evidence that normal SNe~II are constrained to short rise-times and other possible hydrogen-rich SNe have longer rise-times. These SNe were identified based on their exceptionally long rise-times besides their spectra, demonstrating the power of photometric classification from just the early part of the light-curve. If we select out all photometrically identified SNe~II that have long rise-times of a sigma more than 15 days, we find that a 7-10\% (7\% for SDSS-SN and 10\% for SNLS) has a longer pre-maximum behaviour from normal SNe~II. We argue that most of these SNe are actually of a different origin but we keep them in the analysis since we do not have a spectroscopic confirmation. If we were to take them out when calculating the median values shown in the figures and in Table~\ref{table-rise}, we would obtain median Monte Carlo rise-times shorter by $\sim$1 day.

\subsection{Rise-time differences versus effective wavelength}

\begin{figure*}%[htbp]
\centering
\includegraphics[width=0.49\linewidth]{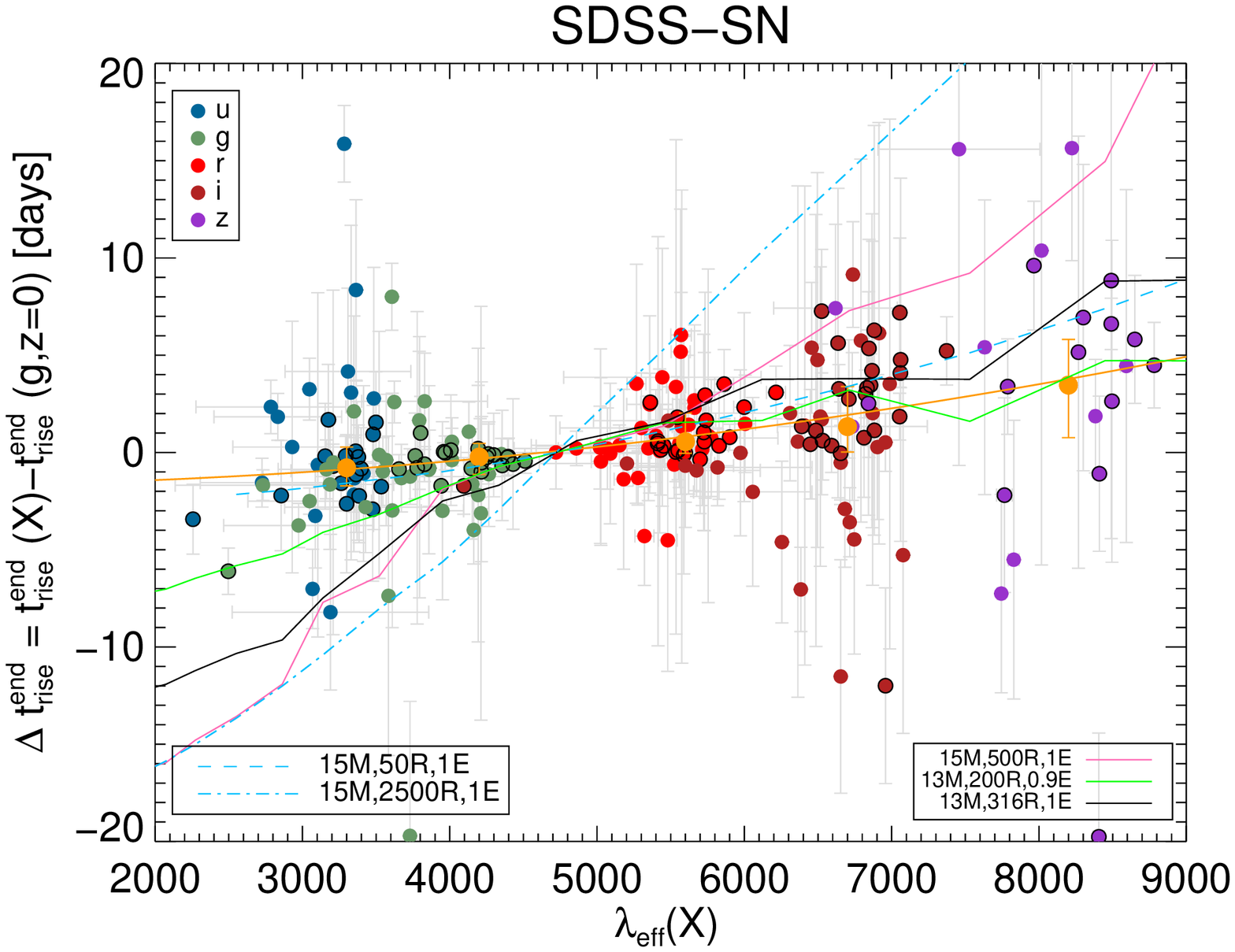}
\includegraphics[width=0.49\linewidth]{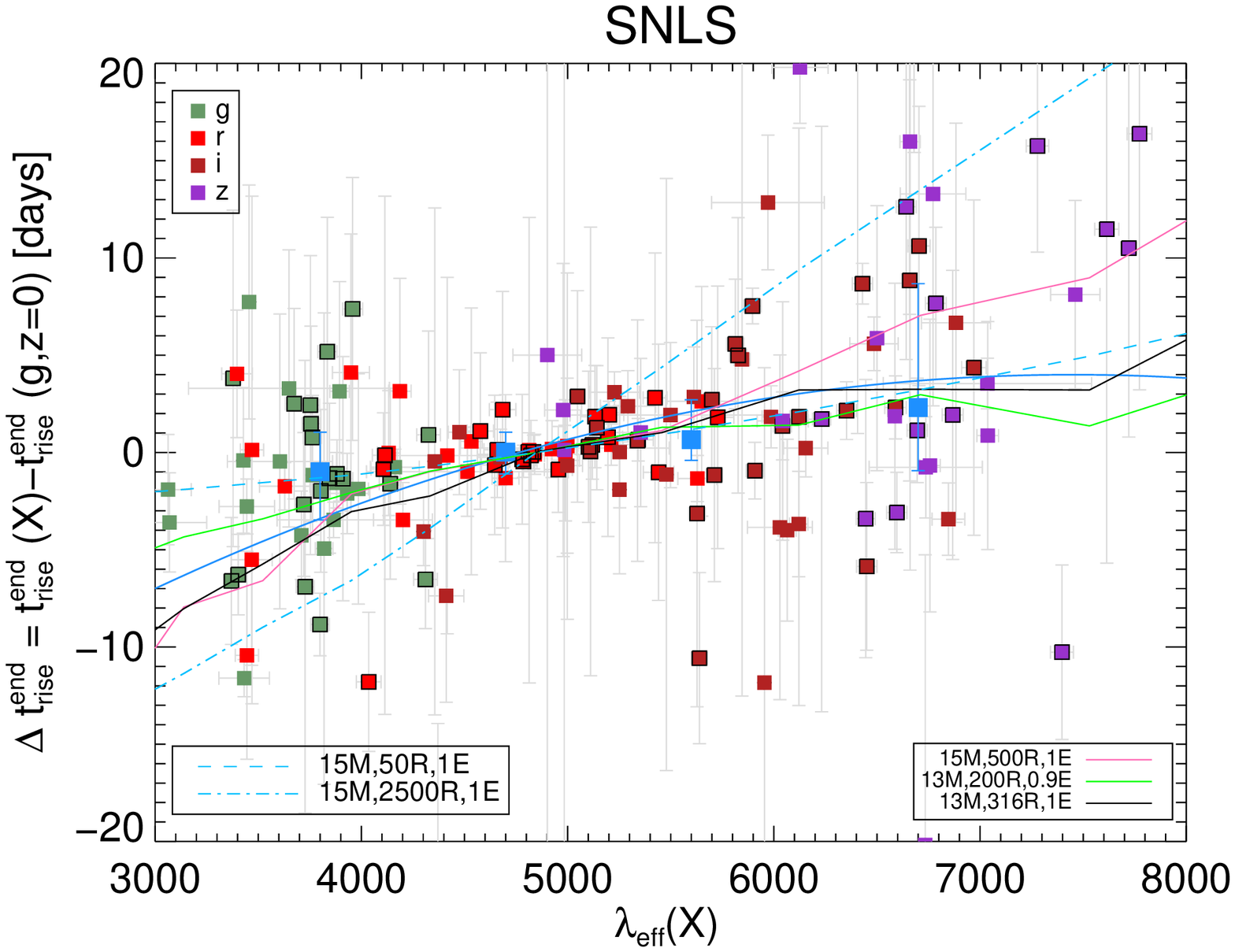}
\caption{Difference in rise-times, $\Delta$\fitrise, respect to an interpolated rise-time at $g$-band rest-frame (\lambdaeff=4722\AA) for the SDSS-SN (left) and SNLS (right) samples. Points with black contours are the golden samples and larger symbols are the median values at characteristic wavelengths. The best second-order polynomial fit is shown as a solid line (orange for SDSS-SN and blue for SNLS). Overplotted are representative analytical models from RW10 for extreme radii of 50\rsun\,(blue dotted) and 2500\rsun\,(blue dashed), and hydrodynamical models from T09 for a standard 500\rsun\,(pink), a more compact 200\rsun\,(green) and a dense progenitor with 316\rsun\,(black).}
\label{deltaexprise-effw}
\end{figure*}

One can see in Figures~\ref{fitrise-effw} and ~\ref{exprise-effw} that the rise-times are shorter at bluer wavelengths, i.e. bluer bands peak before redder bands, which is expected from cooling of the photospheric temperature shifting the peak of the emission towards longer wavelengths as time evolves. The relation can be further seen in the rise-time differences of each SN respect to an interpolated rise-time at rest-frame $g$-band (Figure~\ref{deltaexprise-effw}). This trend is similar to the behaviour observed in other SNe such as stripped-envelope SNe \citep{Taddia14}. We fit a second order polynomial $a\lambda^2+b\lambda+c$ to these differences and find that they represent the data better than a linear trend according to an F-test. Comparing these fits with stripped-envelope SNe, we do not find a statistically significant difference although the physical processes driving the rise are of a different nature: for stripped-envelope SNe the rise-times are powered by radioactive nickel heating whereas SNe~II mostly follow the shock-heated cooling at these epochs. 

Although the models also show an increase in rise-time at longer wavelength, the relation is generally steeper for the analytical models than for the observations, except for very small radii ($\sim~$50 \rsun), typical of BSG progenitors, that are inconsistent with the bulk of the absolute measured rise-times. The set of hydrodynamical models of progenitors with small radii ($\sim200$\rsun), on the other hand, are more consistent with the differences between rise-times at different wavelengths. Regardless, both analytical and hydrodynamical models favor again small radii based on the rise-time differences at varying wavelengths. We have compared the photospheric temperature evolution of models with different radii and find that larger pre-SN radii lead to a slower temperature evolution. The slower temperature evolution results in later peaks at red wavelengths, leading to a steeper dependence of rise-time with wavelength.

\begin{table*}
 \centering
\caption{Median and deviation on the median for the difference between the rise-times, \fitrise, at different wavelengths respect to the effective wavelength at rest-frame $g$-band (4722\AA) for the SDSS-SN and SNLS silver and golden samples.}
\label{table-deltarise}
 \begin{tabular}{|c|c|c|c|c|}%c|c|c|c|}

Effective wavelength &  \multicolumn{2}{c|}{SDSS-SN} &  \multicolumn{2}{c|}{SNLS} \\
(\AA) & silver & golden & silver & golden \\
 \hline
 3300 & $-0.8^{+1.1}_{-0.9}$ &  $-0.7^{+0.8}_{-1.0}$ & -- &  --  \\
 3800 &  -- &  -- & $-1.0^{+2.0}_{-2.5}$ &  $-0.7^{+2.8}_{-2.0}$  \\
 4200 & $-0.2^{+0.7}_{-0.9}$ &  $-0.2^{+0.3}_{-0.6}$ & -- & --  \\
 4700 & -- &  -- & $0.0^{+1.0}_{-1.1}$ &  $0.1^{+1.2}_{-0.9}$  \\
 5600 & $0.6^{+1.5}_{-0.6}$ &  $0.6^{+2.0}_{-0.5}$ & $0.7^{+2.1}_{-1.1}$  &  $1.1^{+1.6}_{-1.5}$  \\ 
 6700 & $1.3^{+2.1}_{-1.3}$ &  $1.8^{+1.7}_{-1.4}$ & $2.3^{+6.4}_{-3.3}$ &  $2.7^{+7.8}_{-1.6}$ \\
 8200 & $3.5^{+2.4}_{-2.7}$ &  $4.1^{+1.7}_{-1.6}$ & -- & -- \\
\end{tabular}
\end{table*}

\subsection{Power-law}

\begin{figure}%[htbp]
\centering
\includegraphics[width=1.\linewidth]{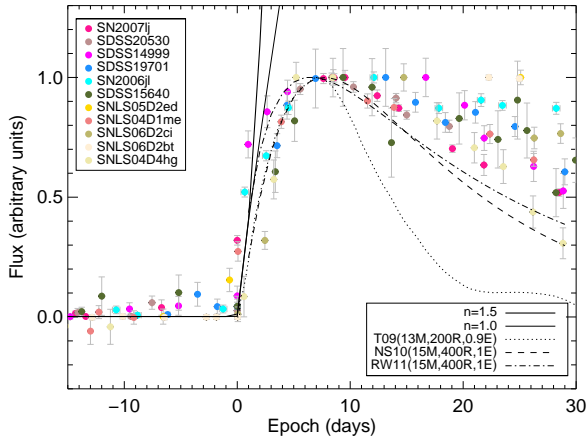}
\caption{Aggregate light-curves of 11 SNe~II from the SDSS-SN and SNLS that have fitted power-law rise-times between 6.5 and 7.5 days. The solid lines shows power-laws of $n=1.5$ (thick) and $n=1.0$ (thin). The dotted line is the hydrodynamical model by T09 with R=200 \rsun\, (with $n=1.3$) while the dashed and dot-dashed are the analytical models by NS10 and RW11 with R=400 \rsun\, (with $n=1.0$ and $n=0.8$ respectively).}
\label{overrise}
\end{figure}

The power-law of the rising light-curves can provide important information on the explosion physics of SNe \citep[e.g.][]{Piro13}. From the fits to equation~\ref{eq_pow} we can obtain the power law index $n$ for SNe with sufficient early coverage. The result of the fit is highly dependent on the range of epochs used in the fit. For simplicity, we take data before half the maximum flux in a given band. As the rise-time of SNe~II is fast, this is only possible for a small subset of SNe~II. We obtain a median power index of $0.96^{+1.10}_{-0.77}$ for the SDSS-SN and $0.91^{+1.42}_{-0.59}$ for the SNLS, where the upper and lower uncertainties represent the one $\sigma$ dispersion of the distribution. Although most of the SNe that we can calculate this for require data in the rise and are therefore biased to longer rise-times, we do not find any evidence for a trend of the power-law with effective wavelength or rise-time duration, and this is confirmed for the analytical and hydrodynamical models. Taking advantage of the combined statistics of multiple light-curves, we can make aggregate light-curves for SNe of a given rise-time duration to calculate the power-law. In figure~\ref{overrise}, we add the light-curves of 11 SNe with rise-times between 6.5 and 7.5 days to a common date of explosion obtained through individual power-law fits. We can see some scatter in the rise which may come from uncertainties in the explosion dates of each SN. Nevertheless, we obtain consistently low power-law indexes of $\lesssim1.4$ for several aggregate light-curves of different rise-time durations between 5 and 15 days.

We note that similar power-law fits to both, analytical and hydrodynamical models, also predict such shallow power-laws. This is not surprising: based on the analytical equations of NS10, \citet{Piro13} show that the UV and optical luminosity of a light-curve dominated by shock-heated cooling rises as $t^{1.5}$. When a light-curve rises more steeply than this, it probably means that other effects such as radioactive heating or recombination have started to play a role in the energy input. Given that most of our individual SNe and all aggregate light-curves have power-law indexes lower than 1.5, this provides strong evidence that the early light-curves of most SNe~II are powered by shock-heated cooling. For all models, we see that the post-maximum slopes decline much faster than for the observations. This will be further discussed in section~\ref{slopes}.

\subsection{Inferred pre-explosion radii}\label{inferradii}

\begin{figure}%[htbp]
\centering
\includegraphics[width=1\linewidth]{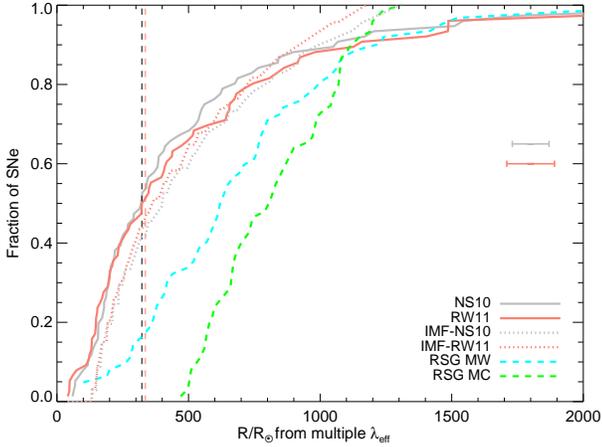}
\caption{Cumulative distribution of inferred radii for the golden samples of SDSS-SN and SNLS together calculated from multi-wavelength rise-times, \fitrise, with models NS10 (solid gray) and RW11 (diagonal red). Vertical dashed lines show the median radii for both models. The dashed cyan line shows the distribution of RSG radii from the MW \citep{Levesque05} and the dashed green line the distribution of RSG radii from the MC \citep{Levesque06}. Dotted lines are Salpeter IMF distributions of expected radii assuming $R\sim1.4M^{2.2}$ with initial masses of 8-30\msun. Median errors on the radii are shown as horizontal bars in the right.}

\label{expradii}
\end{figure}

If the dispersion in the rise-times found in previous sections is real, the most plausible progenitor property in the analytical models that can account for this is radius (see Figure~\ref{exprise-effw}). Pursuing this assumption further, one can then use NS10 and RW11 to infer radii based solely on the rise-time duration and keeping the other parameters fixed at 15 \msun\, and 1 foe. Taking the multiple fitted rise-times at different wavelengths of each SN weighted by their error, we can then compare to the values measured for analytical models and obtain an optimal radius. The cumulative distribution of these radii for the golden samples of both SDSS-SN and SNLS together are shown in Figure~\ref{expradii} for the two models. The distributions predicted by NS10 and RW11 are consistent (Kolmogorov-Smirnov KS-test probability of being drawn from the same population of 0.53) and the median radii are remarkably similar: $323^{+442}_{-166}$ and $336^{+525}_{-199}$\rsun\, ($385^{+603}_{-210}$ and $418^{+986}_{-253}$\rsun\, for the silver samples). The uncertainties here quoted represent the 84\% dispersion range. If we perform a Monte Carlo simulation in which we measure new radii from simulated rise-times based on the uncertainties and covariances, we find consistent mean radii of all simulations of $338^{+407}_{-202}$ and $347^{+590}_{-280}$ \rsun\, ($395^{+610}_{-245}$ and $425^{+1248}_{-266}$\rsun\, for the silver samples) for each analytical model respectively.

We find that less than 5\% of the objects are consistent with BSG progenitors powered exclusively by shock cooling of radii less than 100\rsun, although we are limited by the cadence of the surveys. This is roughly in agreement with the 2\% fraction inferred for SN~1987A events \citep{Kleiser11}. On the other hand, some objects having long rise-times have radii of several thousands \rsun. As explained in section~\ref{longrise} these objects probably have a different powering source. If we do not take into account the identified peculiar SNe with long rise-times, we also find less than 10\% of our sample having longer radii than 1200\rsun. These radii are also confirmed qualitatively with the hydrodynamical models of T09 (see appendix~\ref{lte-stella} for non-LTE treatment). In Figure~\ref{fitrise-effw} one can see that their models of radii much larger than 500\rsun\, are ruled out and in fact smaller radii of $\sim100-400$\rsun\, are favoured. Small radii have been recently inferred through modelling of individual subluminous SNe~II like SN~2008in \citep[126\rsun,][]{Roy11} and SN~2009N \citep[287\rsun,][]{Takats14}.

\begin{figure}%[htbp]
\centering
\includegraphics[width=1\linewidth]{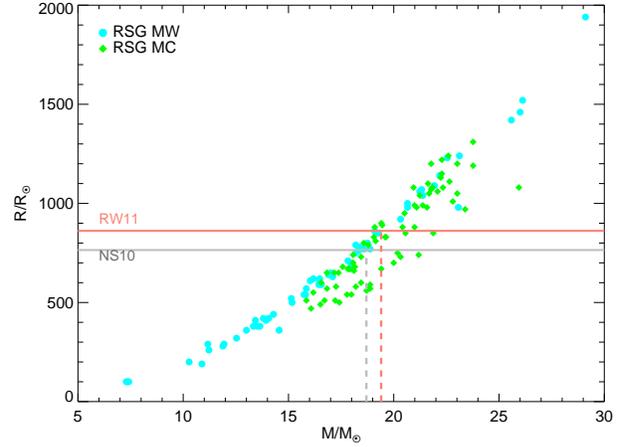}
\caption{Radius versus mass for RSG in the MW and MC \citep{Levesque05,Levesque06} and 84\% upper limits of our radius distribution with NS10 and RW11 models as solid lines. Dashed lines are the approximate mass correspondance for the MW.}

\label{radius-mass}
\end{figure}

Typical RSG span radii of 100-1600\rsun\, \citep{Levesque05,Arroyo13} and interferometric studies confirm the existence of some very extended nearby stars with more than 900\rsun\, \citep{Haubois09,Wittkowski12}. According to our findings, most SNe~II are produced by RSGs with radii at the lower end of this distribution. This is shown in Figure~\ref{expradii} where we present the distribution of RSG radii for the Milky Way (MW) and Magellanic Clouds (MC) \citep{Levesque05,Levesque06}. These were obtained with Stefan Boltzman's law from the effective temperatures and bolometric luminosities inferred with MARCS stellar atmosphere models \citep[see][]{Plez92,Gustafsson75} applied on observed spectrophotometric data. The distributions are inconsistent with the radii inferred from our SN rise-time analysis (KS-test of being drawn from the same population of less than $10^{-5}$ and $10^{-15}$ for the MW and MC respectively). Since smaller stars are also less massive, one is tempted to associate this observation with the seemingly lack of massive RSG progenitor of more than $\sim18$ \msun\, in pre-explosion images \citep{Smartt09}. Figure~\ref{radius-mass} shows RSG radii against the masses obtained from the mass-luminosity relation for the MW and MC \citep{Levesque05,Levesque06}. We include our one $\sigma$ (84\%) upper bounds on radii for NS10 and RW11 models. They roughly correspond to upper limits of 19 \msun, consistent with the limits from pre-explosions masses. This would be an independent confirmation of their result. 

Nevertheless, the disagreement between our inferred radii and the observed RSG from the MW could just be a result of the incompleteness of their survey for which the smaller RSG are not all observed and are therefore missing in the distributions. Instead, we can do a very basic comparison with the expectations of a standard Salpeter Initial Mass Function (IMF) assuming all SNe~II arise from massive single stars between 8 and 30\msun\, (the upper limit has very little impact here). To obtain radii, we use a mass-radius relation of $R/R_{\odot}=1.4(M/M_{\odot})^{2.2}$ from a log-log linear fit to the MW RSG of \citet{Levesque06}. The resulting simplistic radius distributions of this IMF are shown in dotted lines in Figure~\ref{expradii}. The normalization is obtained by fitting the IMF to the NS10 and RW11 distributions. In this case, our inferred distributions are more consistent with the simple IMF distributions, especially for RW11 (KS-test of being drawn from the same population of 0.03-0.09 for RW11 and 0.09-0.2 for NS10).

On the other hand, another possible reconciliation stems from recent RSG observations extending to the IR which suggest that they are hotter than previously thought \citep{Davies13}, which can result in a reduction by up to 30\% of the inferred RSG radii \citep{Dessart13}. Arbitrarily shifting the distributions of RSG radii in the MW by this amount results in a population that is more consistent with our inferred radii (KS-test of being drawn from the same population of $0.02-0.1$). Furthermore, there are many uncertainties regarding the stellar models used to obtain these radii, in particular the mixing length can greatly affect the inferred radius \citep[e.g.][]{Deng00,Meynet14}. These difficulties raise the question if RSG radii are systematically over-estimated and the distribution is more similar to what we find based on SN rise-times.

Finally, metallicity can affect the rise-times through changes in the pre-SN progenitor structure \citep{Langer91,Chieffi03,Tominaga11}. Lower metallicity results in shorter radii: according to models by \citet{Chieffi03} a reduction from solar metallicity $Z=0.02$ to $Z=0.001$ reduces the pre-SN radius by $\Delta R=150-600$\rsun\, for initial masses of 13-15\msun. This could explain our findings but it would mean that most SN~II are hosted in low metallicity environments which is not a general observation for nearby targeted hosts \citep[e.g.][]{Anderson10} but should be tested for our sample. The distribution of RSGs in the low-metallicity MC does not support this either, nor the longer rise-times found for the SNLS, which at higher redshift is expected to have lower mean metallicity.

\begin{figure*}%[htbp]
\centering
\includegraphics[width=0.49\linewidth]{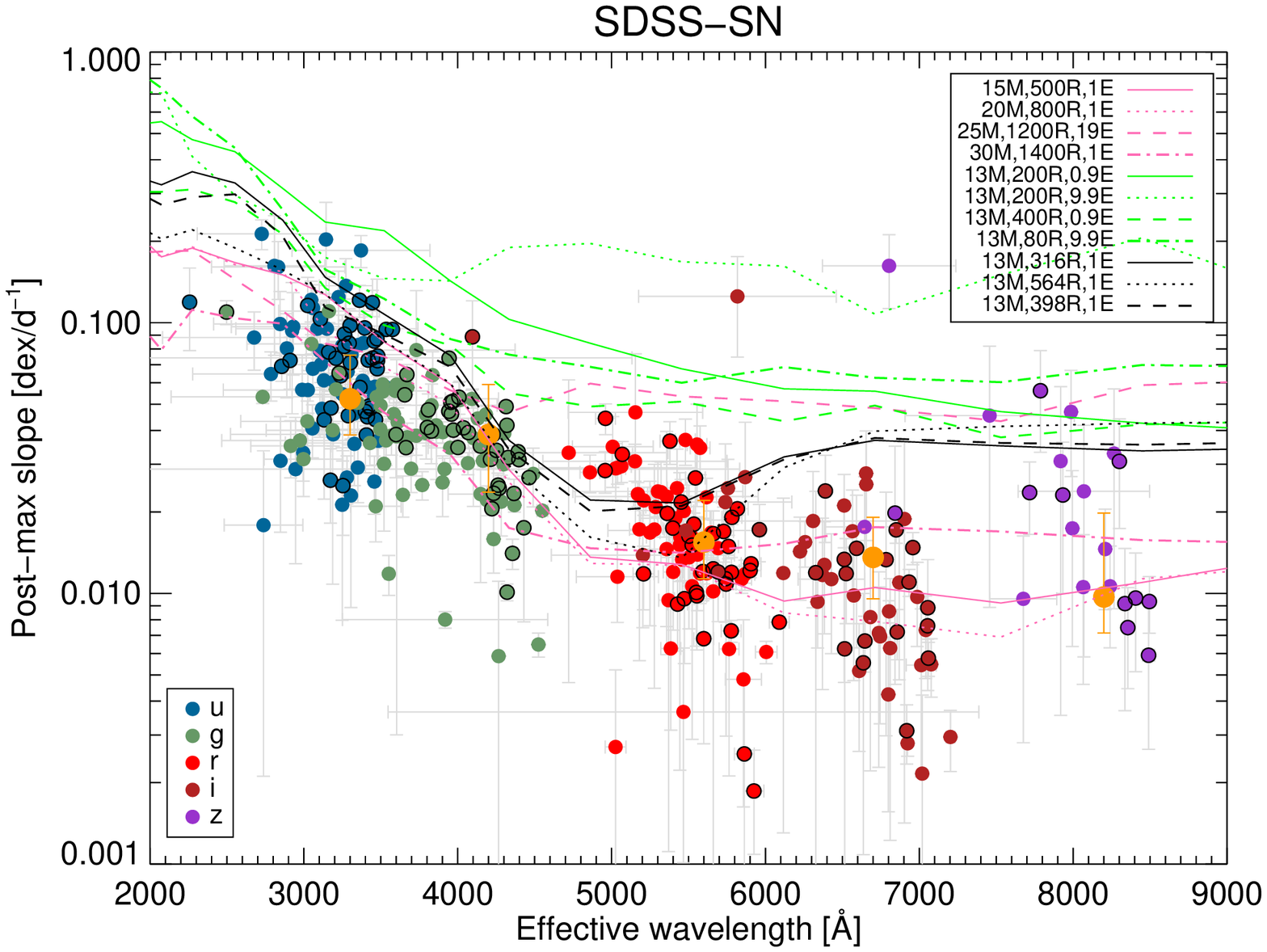}
\includegraphics[width=0.49\linewidth]{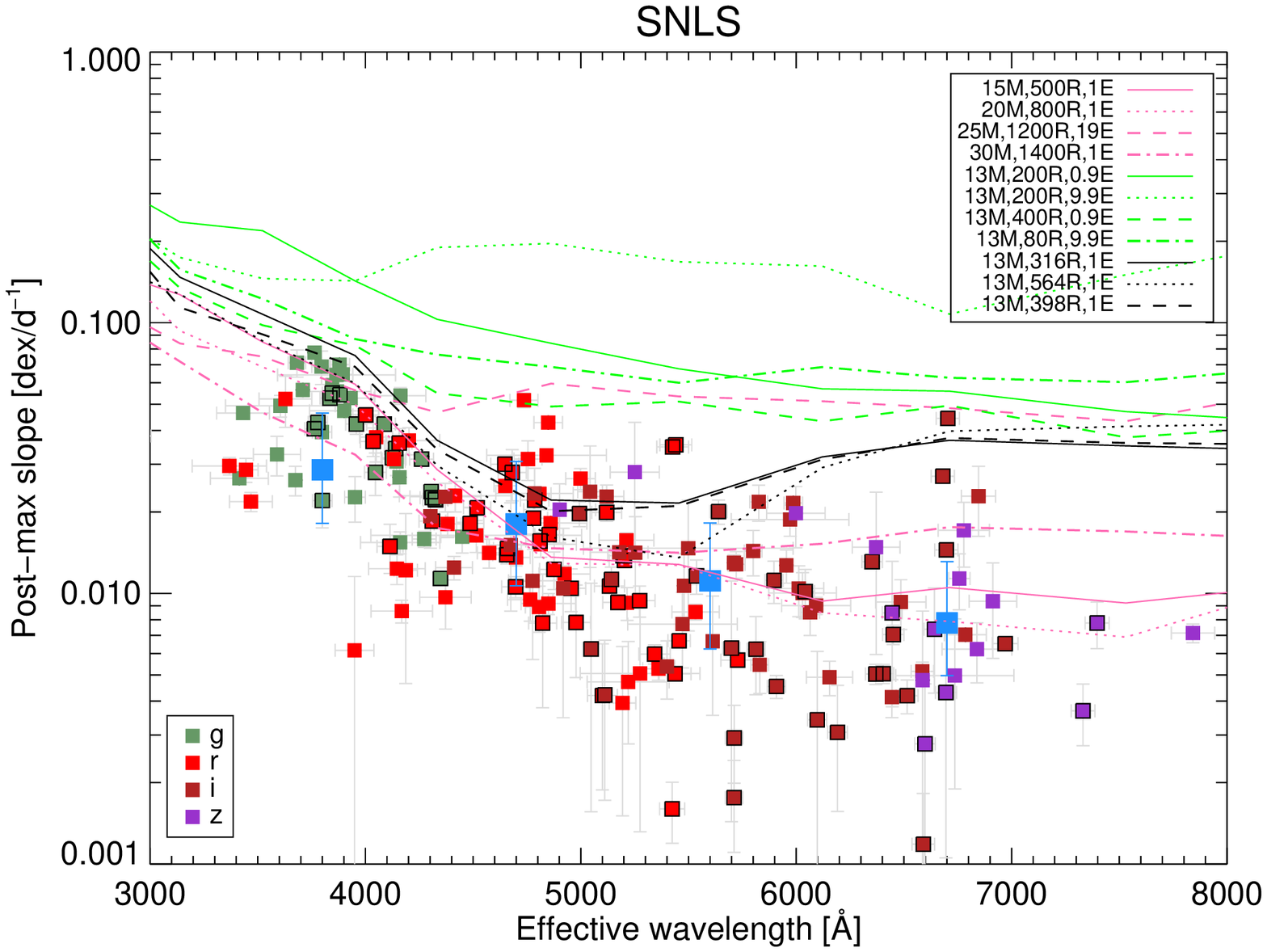}
\caption{Post-maximum slopes as a function of effective wavelength for the SDSS-SN (left) and SNLS (right) silver (all points) and golden (points with black contours). Standard (pink), small-radii (green) and dense (black) hydrodynamical models from T09 are shown in lines. We do not present analytical models because NS10 and the simple model of RW11 that we use do not treat the recombination of hydrogen.}
\label{slope-effw}
\end{figure*}

\subsection{Slopes versus effective wavelength}\label{slopes}

Here we analyze the post-maximum linear slopes measured in each band as a function of effective wavelength for both samples (see figure~\ref{slope-effw}). At the shortest wavelengths, the light-curves decay much faster than at longer wavelengths, a consequence of the temperature evolution of SNe~II. The decay rate of optical wavelengths above approximately 5000\AA\, are more similar to each other reflecting the known hydrogen recombination wave as it recedes through the ejecta. The slope and duration of this plateau phase are possibly affected by the mass of hydrogen present in the ejecta. This can be best seen in the hydrodynamical models of T09: models with large radii (pink) predict the right range of slopes whereas the set of models with small radii (green) overpredict the slopes. This result can be explained by a different ejecta mass or radius in the RSG progenitor \citep{Litvinova83,Popov93}. RSG with large radii have a larger hydrogen mass that produces a shallower (and possibly longer) plateau and RSG with small radii have much less hydrogen and hence the recombination wave recedes too fast and does not produce a plateau. Comparing our slopes with the results of \citet{Anderson14}, we obtain a very consistent range of $0.3-4$ magnitudes per 100 days in the wavelength range of 5000-6000\AA\, (roughly equivalent to $V$-band), indicating that the populations at high-$z$ are quite similar to low-$z$. We note that our slopes are a combination of their $s_1$ and $s_2$ values.

Together with the results presented in the previous sections, we then require RSG progenitors of small radii to produce short rise-times but with large hydrogen mass to produce shallow post-maximum slopes. Thus a dense hydrogen envelope is needed which can be achieved through stronger convection in RSG envelopes that leads to more mixing. Indeed, there is discussion on the variation of convective mixing length in different stars \citep{Ferraro06} which is poorly constrained for massive stars \citep{Meakin07}. Varying the mixing length severely affects the radius \citep{Maeder87}: \citet{Dessart13} showed that varying it from $\alpha=1.1$ to 3 in MESA STAR models \citep{Paxton11} changes the radius from 1100 to 500 \rsun, and it does not affect the He core characteristics since the envelopes are only influenced at the end of their lives. They also find that a small radius is needed to solve SN color evolution discrepancies of SN~1999em from stellar evolutionary models coupled with SN radiative transfer. This opens up an interesting test that can be performed in the future, both on the current and other samples, by looking at the color evolution and its relation to rise-time. Also, different treatments of the mixing length theory for certain RSG envelopes in which convective speed becomes supersonic, predict smaller radii for the same mass \citep{Deng00,Josselin07}. We note that enhanced rotation or core overshooting may also alter the radius but in the opposite direction to more extended RSG \citep{Maeder00,Dessart13}.

\begin{figure}%[htbp]
\centering
\includegraphics[width=1\linewidth]{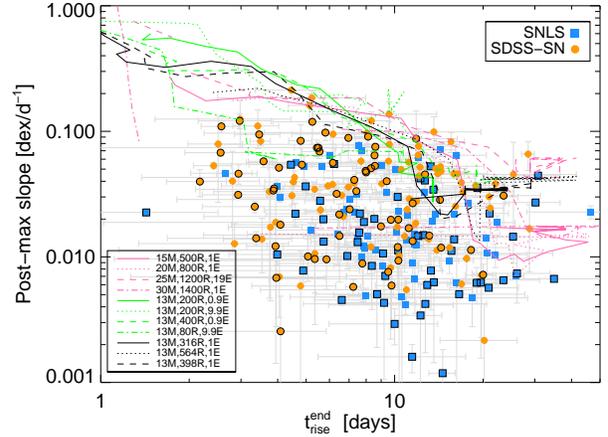}
\caption{Post-maximum slopes versus rise-time for the silver samples of SDSS-SN (orange circles) and SNLS (blue squares). Golden samples have black contours. Measurements in all photometric bands are included. Selected standard (pink), small-radii (green) and dense (black) hydrodynamical models from T09 are shown in lines.}
\label{slope-rise}
\end{figure}

We have included three additional hydrodynamical models with dense envelopes to test this hypothesis further. These models are shown as black lines in previous Figures. In Figure~\ref{slope-rise}, we directly compare the pre-maximum and post-maximum behaviors, i.e. rise-time and plateau decline, for both samples in light of different hydrodynamical models. We do not see any bimodality indicative of two groups, SNe~IIP and SNe~IIL. There is however a trend for slower decliners to have longer rise-times and faster decliners to rise quicker (Spearman correlation of -0.39 and -0.27 for golden and silver samples). Dividing the population by the median rise-time (8.9 days), the distributions of slopes of SNe~II with short rise-times is significantly different from the one with longer rise-times (KS-test of 0.002 for golden and silver samples). In the traditional nomenclature, this means that SNe~IIL tend to have steeper post-maximum slopes and shorter rise-times. \citet{Gall15} find a similar trend. The most straightforward interpretation based on the previous discussion is that SNe~II that decline more rapidly during the plateau have smaller radii. If a smaller radius means less hydrogen, then this is naturally explained. 

The hydrodynamical models indeed predict a relation between the rise-time duration and the post-maximum decline. However, none of them is able to reproduce the parameter space in Figure~\ref{slope-rise}. This leads us to suggest an alternative scenario. \citet[][Figure 7]{Moriya11} and \citet[][Figure 1]{Baklanov13} present models for a family of SNe~IIn, i.e. SNe~IIP with circumstellar material (CSM) interaction. From their figures one can clearly see that the optical rise-times of their models are much shorter than pure SN~IIP models; in fact they can often have values below 10 days as we find in this study. In those models the shock breakout occurs at larger distances, beyond the stellar surface, and the signal is therefore smeared in the CSM and its duration is prolonged. Consequently, the rise-time directly corresponds to shock breakout and not shock cooling. Would it then be possible that normal SNe~II indeed have pre-SN mass-loss high enough to affect the shock and the rise-time but not so strong as to create narrow emission lines in the post-maximum spectra? \citet{Gezari15} recently confirmed a much quicker rise in NUV for SN~IIP PS1-13arp and propose a similar explanation. If this were the case, then faster decliners (SNe~IIL), that seem to have shorter rise-times, may have a larger or denser CSM interaction. The interaction would naturally explain their higher brightness as the kinetic energy is more efficiently converted to thermal energy in the CSM \citep{Moriya11}. In fact, \citet{Valenti15} recently detect moderate CSM interaction for type IIL SN~2013by and suggest such a common feature for all SNe~IIL. Moreover, the recent work by \citet{Smith15} shows the existence of transitional objects between classical SNe~II and interacting SNe~IIn such as PTF11iqb. These objects initially behave like SNe~IIn with narrow emission lines that later disappear resembling more SNe~II. This is further indication that SNe~II and SNe~IIn may all be part of a continuous family with different CSM quantity or density in following increasing order: IIP-IIL-IIn. We note that in principle any SN type can be hidden behind a thick CSM, making it look like a SN~IIn. \citet{Ofek14} actually find a large diversity of rise-times for SNe~IIn. Based on our findings, it is tempting to argue that only SNe~IIn with short rise-times are SNe~II inside a thick CSM, where the energy input is more strongly attributed to the shock breakout.

Finally, a different observation of interest by \citet{Anderson14} in which SN~II $V$-band light-curves (but also observable in others bands such as $B$) are found to be described best with two declining slopes after maximum, $s_1$ and $s_2$, suggests that the first steeper decliner stems from extended cooling, which is related to the radius of the progenitor \citep{Bersten12}. Such late transition times between $s_1$ and $s_2$ require very large radii which, together with the fast rise-times we measure, can only be reconciled in this CSM scenario.

All these observations point towards an important paradigm shift in which SNe~II (IIP and IIL) also have pre-SN wind mass-loss of $10^{-4}-10^{-3}$\msun$yr^{-1}$ lower than SNe~IIn of $0.01-0.1$\msun$yr^{-1}$. The obvious question arises on the coincidence of mass-loss and SN explosion. Either episodic mass-loss is extremely common or there is some mechanism that makes them coincide. \citet{Mcley14,Soker13,Quataert12} present such a possibility: the RSG envelope expands shortly before explosion due to powerful nuclear burning and so called $p$-waves carrying nuclear energy from the core to the outer layers through core convection. This can lead to extended envelopes \citep{Mcley14,Soker13} or mass-loss \citep{Quataert12}, which in both cases results in shorter rise-times. Binary interaction could play an important role in recurrent mass-loss episodes and asymmetric CSM, as observed for some interacting SNe \citep{Mauerhan12,Smith15}.

Instead of CSM, a very extended envelope can also produce short rise durations. As a matter of fact, extended molecular atmospheres have been observed for RSG \citep{Wittkowski12,Arroyo13,Arroyo15} and densities in the outer layers of RSG are larger than convection predicts \citep{Ohnaka11,Ohnaka13}. These explanations naturally fit the short rise-times, shallow plateaus and late transition of $s_1$ to $s_2$.

\begin{table*}
 \centering
\caption{Median and deviation on the median for post-maximum slopes of the SDSS-SN and SNLS silver and golden samples at different effective wavelegnths.}\label{table-slope}
 \begin{tabular}{|c|c|c|c|c|}%c|c|c|c|}

Effective wavelength &  \multicolumn{2}{c|}{SDSS-SN} &  \multicolumn{2}{c|}{SNLS} \\
(\AA) & silver & golden & silver & golden \\
 \hline
 3300 &  $0.052^{+0.024}_{-0.014}$ &  $0.064^{+0.020}_{-0.019}$ & -- & -- \\
 3800 &  -- &  -- &  $0.043^{+0.011}_{-0.021}$ & $0.028^{+0.014}_{-0.007}$ \\
 4200 &  $0.039^{+0.021}_{-0.015}$ &  $0.044^{+0.027}_{-0.018}$ & -- & -- \\
 4700 &  -- &  -- &  $0.018^{+0.013}_{-0.007}$ &  $0.019^{+0.013}_{-0.008}$ \\
 5600 &  $0.016^{+0.007}_{-0.004}$  &  $0.012^{+0.006}_{-0.003}$ &  $0.011^{+0.008}_{-0.005}$ & $0.009^{+0.005}_{-0.004}$ \\ 
 6700 &  $0.014^{+0.006}_{-0.004}$ &  $0.012^{+0.005}_{-0.003}$ & $0.008^{+0.005}_{-0.003}$ & $0.006^{+0.004}_{-0.003}$ \\
 8200 &  $0.010^{+0.010}_{-0.003}$ &  $0.010^{+0.010}_{-0.002}$ & $0.007^{+0.010}_{-0.002}$ &  $0.007^{+0.038}_{-0.004}$ \\
\end{tabular}
\end{table*}

\section{Summary}\label{summary}
We have measured the rise-times as a function of effective wavelength for a large sample of SNe~II spanning a wide redshift range of $0<z\lesssim0.7$ from the rolling surveys of SDSS-SN and SNLS. The rise-times increase with wavelength as expected for shock-heated cooling. The power-law indexes of the rise are generally lower than 1.5, also confirming the expected behaviour of cooling of shocked material \citep{Piro13}.

At an effective wavelength of 4722\AA, the characteristic $g$-band rest-frame wavelength, we find a combined average rise-time duration of $(7.5\pm0.3)$ days. The distribution of rise-times spans a one $\sigma$ range of $4-17$ days. We caution the reader that given the cadences of our surveys, there could be more SNe~II with even shorter rise-times. From this continuous distribution, we find no evidence for two populations of SNe~IIP and SNe~IIL. We do identify several SNe~IIn and peculiar SN~1987A-like SNe~II based solely on the long duration of their rise. This opens up interesting capabilities for the early photometric identification of SNe~II and other SNe in current and future wide-field transient searches such as DES \citep{Bernstein14} and LSST \citep{Ivezic11}.

We compare our findings to predictions from analytical models by NS10 and RW11, as well as hydrodynamical models from T09. The rise-times in the models are most sensitive to radius. The measured rise-times we obtain are much shorter than typical progenitor models of RSG with radii larger than 500 \rsun. Using the analytical models and a new set of hydrodynamical models with smaller radii, we find that the bulk of our SNe are consistent with RSG progenitors of 200-500 \rsun. The distribution of rise-times that we obtain translates into 84\% of radii lying between 100-900 \rsun. The mean radii we obtain are inconsistent with typical expectations of RSG \citep{Levesque05,Neilson11}. In order to double the size of the radius to $\sim800$\rsun\, in the analytical models, one would need to systematically increase the rise-time by 7-8 days, which is well beyond our systematic uncertainties. These smaller radii may confirm that previous RSG temperatures were under-estimated \citep{Davies13} or perhaps that mixing length theory is inaccurate \citep{Deng00}.

From the shallow post-maximum slopes that we measure for our sample compared to hydrodynamical models, we infer that large envelope masses of hydrogen are needed, as has been previously known for SNe~II \citep{Popov93}. However, given the small radii found through the rise-time calculations, RSG progenitors with dense envelopes are required. One way to achieve this is through larger mixing length parameters in RSG progenitors \citep{Maeder87}. We include new hydrodynamical models of exploding RSG with dense envelopes that are still not capable of explaining our results, but there is evidently a large range of pre-SN models that are not considered here.

Alternatively, the fast rise-times and long plateaus could be simultaneously explained if RSG have more extended envelopes or denser outer layers \citep{Wittkowski12,Arroyo15,Ohnaka13}, or suffer mass-loss prior to explosion, as has also been recently suggested \citep{Gezari15,Valenti15,Smith15}. The very early rising light-curve in this case would not be due to shock-heated cooling but would be the delayed prolonged shock breakout emission of typical interacting SNe \citep{Moriya11}. This hypothesis will need to be tested with corresponding theoretical models.

Such interpretation brings up the question of the difficulty to simultaenously detect the RSG shock breakout precursor prior to the SN rise and the rise itself. Previous claims of shock breakout detections \citep{Gezari08,Schawinski08,Gezari15} were not accompanied with a clear separation from the main light-curve. This could be an observational bias or could signify that such precursor actually does not exist because it is smeared into the main light-curve. Modern high-cadence searches such as Kepler \citep{Olling14} or HiTS \citep{HiTS-ATEL1,HiTS-ATEL2} will be able to elucidate this question with higher precision than ever before.

\section*{Acknowledgements}
We thank T.~Jos\'e Moriya for useful discussions. This work is based on observations of the Supernova Legacy Survey (SNLS) obtained with MegaPrime/MegaCam, a joint project of CFHT and CEA/DAPNIA, at the Canada-France-Hawaii Telescope (CFHT) which is operated by the National Research Council (NRC) of Canada, the Institut National des Science de l'Univers of the Centre National de la Recherche Scientifique (CNRS) of France, and the University of Hawaii.

Funding for the creation and distribution of the SDSS and SDSS-II has been provided by the
Alfred P. Sloan Foundation, the Participating Institutions, the National Science Foundation, the U.S. Department of Energy, the National Aeronautics and Space Administration, the Japanese Monbukagakusho, the Max Planck Society, and the Higher Education Funding Council for England. The SDSS Web site is \url{http://www.sdss.org/}. The SDSS is managed by the Astrophysical Research Consortium for the Participating Institutions. The Participating Institutions are the American Museum of Natural History, Astrophysical Institute Potsdam, University of Basel, Cambridge University, Case Western Reserve University, University of Chicago, Drexel University, Fermilab, the Institute for Advanced Study, the Japan Participation Group, Johns Hopkins University, the Joint Institute for Nuclear Astrophysics, the Kavli Institute for Particle Astrophysics and Cosmology, the Korean Scientist Group, the Chinese Academy of Sciences (LAMOST), Los Alamos National Laboratory, the Max-Planck-Institute for Astronomy (MPIA), the Max-Planck-Institute for Astrophysics (MPA), New Mexico State University, Ohio State University, University of Pittsburgh, University of Portsmouth, Princeton University, the United States Naval Observatory, and the University of Washington.

S.G., L.G. and F.B. acknowledge support from CONICYT through FONDECYT grants 3130680, 3140566 and 3120227. Support for S.G., L.G., F.B., C.G., F.F., G.P., T.dJ. and M.H. is provided by the Ministry of Economy, Development, and Tourism's Millennium Science Initiative through grant IC120009, awarded to The Millennium Institute of Astrophysics, MAS. MS acknowledges support from the Royal Society. P.B. is supported by the Swiss National Science Foundation grant IZ73Z0\_152485 SCOPES and S.B. by the Russian Science Foundation Grant No. 14-12-00203.

\bibliographystyle{mn2e}
\bibliography{astro}

\appendix

\section{Non-LTE effects on the rise}\label{lte-stella}
\begin{figure*}%[htb]
  \centering
\includegraphics[width=0.49\linewidth]{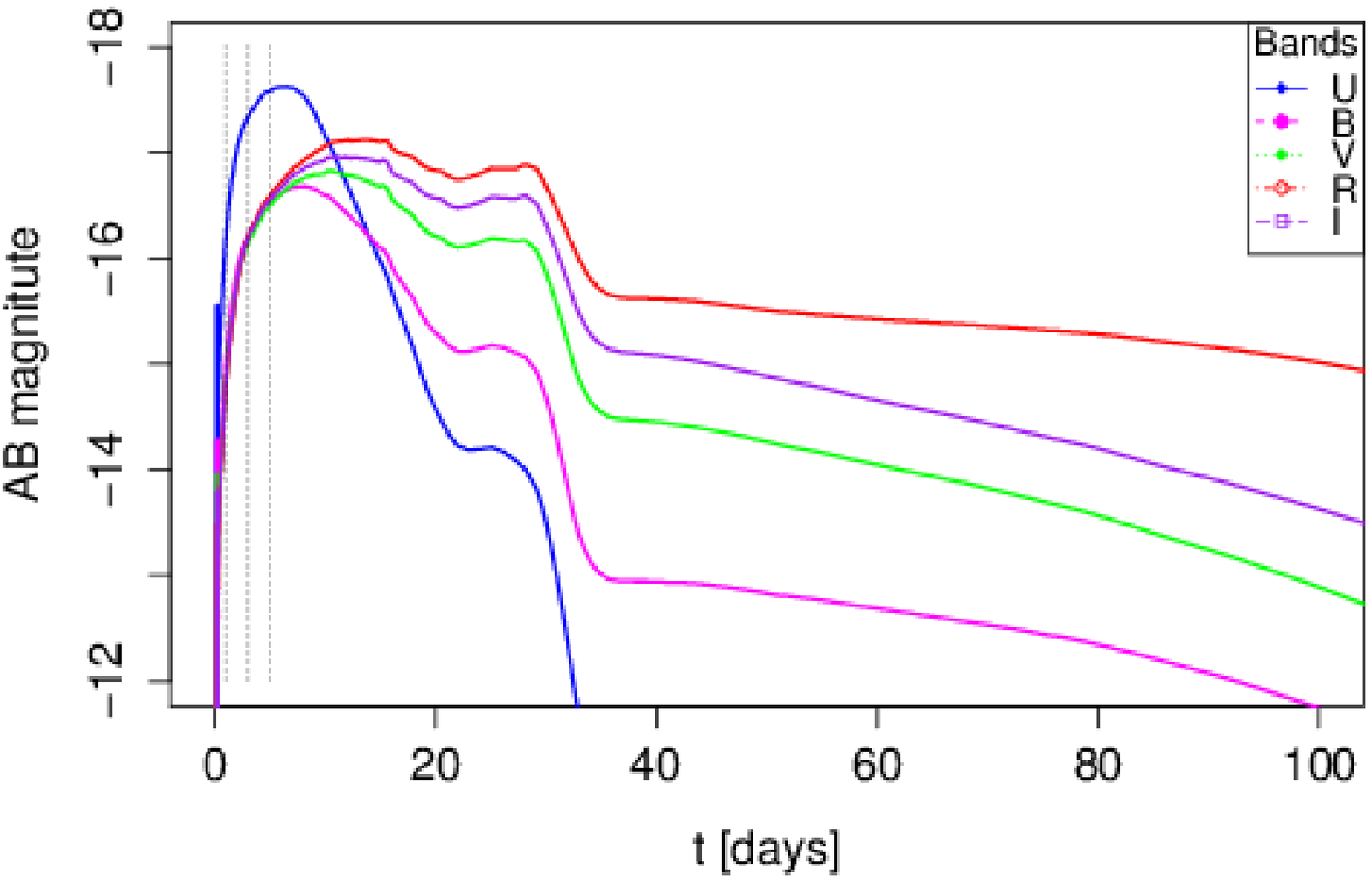}  
\includegraphics[width=0.49\linewidth]{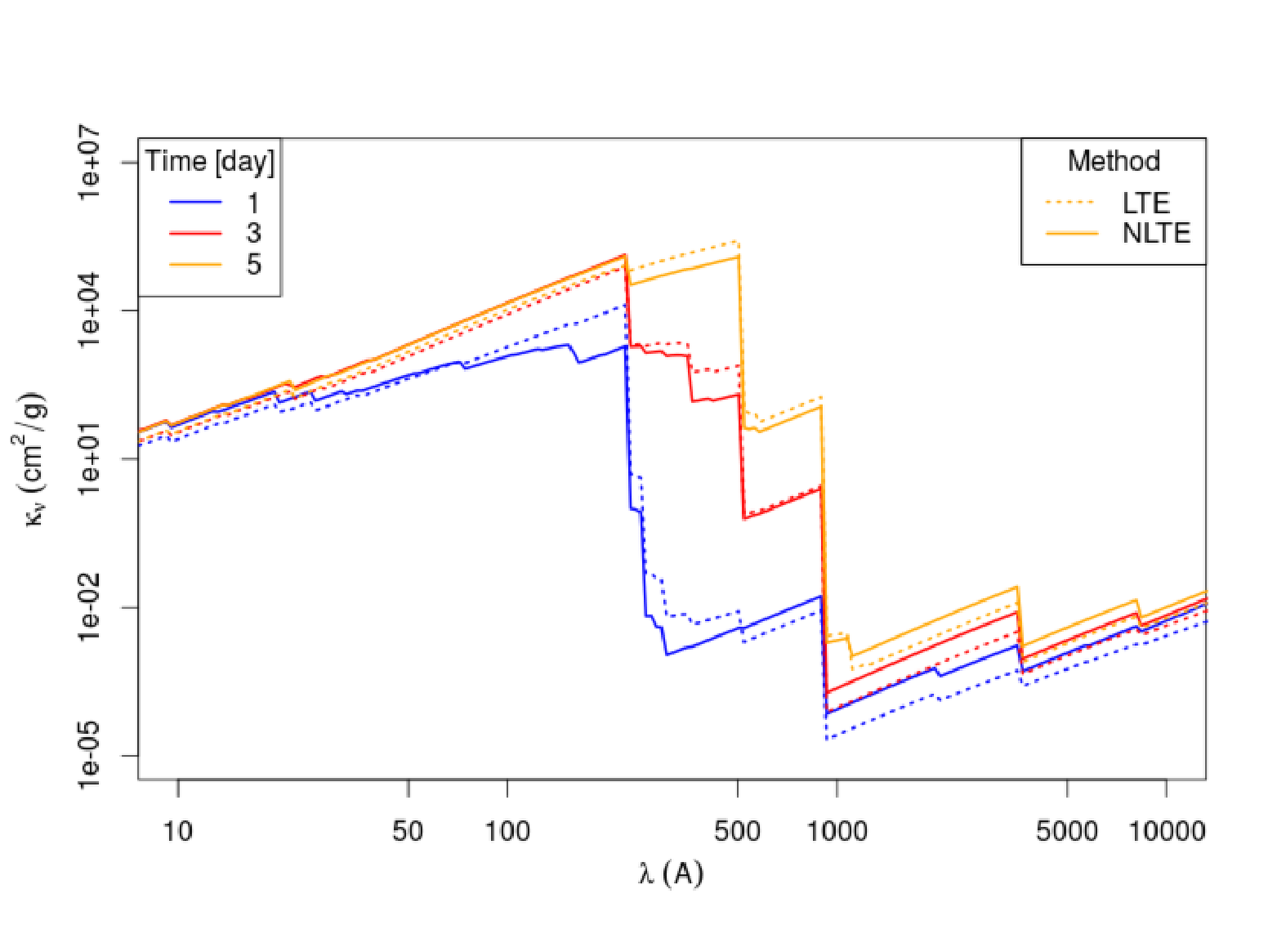}
  \caption{Left: $UBVRI$ light curves of SN models. The dotted lines show three time points $ t = 1,3,5 $~days used for the analysis of the role of the non-LTE effects. Right: The wavelength-dependent expansion opacities, $\kappa$, are plotted at time moments $ t = 1,3,5 $~days. The opacity is calculated using both LTE (dotted) and non-LTE (solid) assumptions to demonstrate the reliability of the LTE approach.} \label{nonlte}
\end{figure*}

The numerical model of T09 was computed with the code {\sc Stella} assuming Local Thermodynamical Equilibrium (LTE) conditions. Under this assumption, opacity, which determines the radiative transfer through a supernova envelope, depends only on local medium properties such as density, chemical composition and temperature. Although the temperature of electrons was calculated in self-consistent simulations with the non-equilibrium spectrum of radiation, in reality the state of excitations and ionizations depends not only on this temperature but also directly on the details of the non-equilibrium spectrum of radiation. In the inner envelope layers the radiation is trapped and the LTE assumption is fulfilled. However, at the outer edge of the envelope, where the radiation is free to leave a supernova, the LTE assumption is violated. Thus the equation of state must be solved taking into account that the radiation spectrum is not in equilibrium with the matter \citep{Mihalas78}. One needs to check how much the non-LTE effects will affect the shape of the light curve and consequently the rise time.

To evaluate this effect, a model from the set of small radii by T09 was chosen. On the rising part of the light curve, three points were chosen at times $ t = 1,3,5 $~days (see left Fig.~\ref{nonlte}). At these epochs the opacity was computed taking into account the non-equilibrium radiation (for details, see \citealt{Baklanov13}). The LTE and non-LTE opacities are shown in right Fig.~\ref{nonlte}. In the range of optical bands the non-LTE opacity increases comparatively to the LTE case due to increased Thomson scattering. This growth is associated with a higher level of ionization of the medium in the field of non-equilibrium radiation, for which color temperature is higher than the local electron temperature in this volume.

Thus at the rising stage of the light curve the difference between the two approaches is small and  for the integral light curves its effect is even smaller. In addition, the Thomson scattering is independent of frequency, and therefore it does not change the shape of the light curves. Therefore, these calculations show that within the accuracy of observations, the use of the LTE models is correct for measurements of the rise time.

\section{Systematic effects}\label{syst}

\subsection{SNII spectral template}
\begin{figure}%[htbp]
\centering
\includegraphics[width=0.99\linewidth]{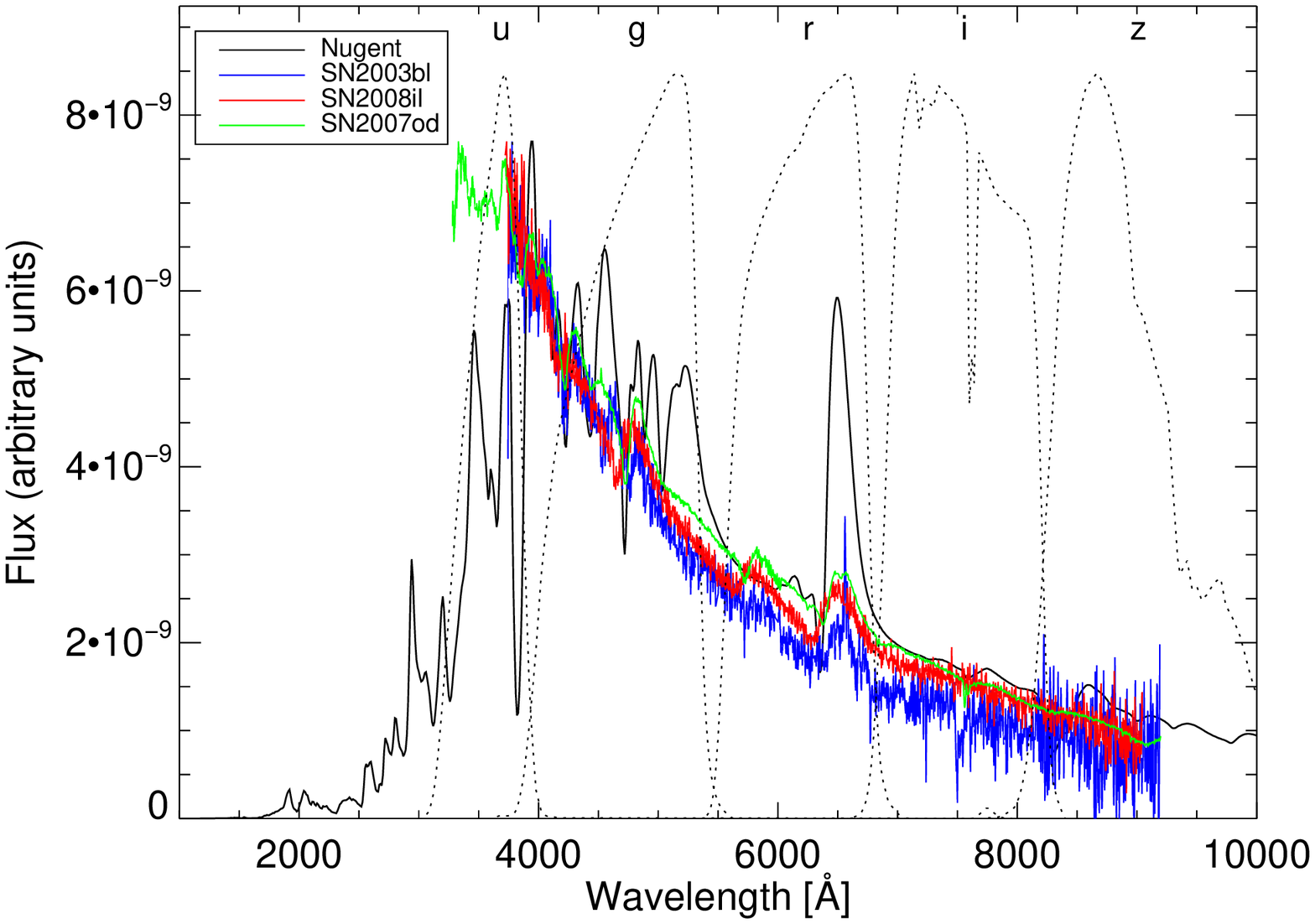}
\caption{SN~II spectral template by Nugent (black) compared to observed rest-frame MW-corrected SN~II spectra of SN~2003bl at $\sim$2 days after explosion (blue), SN~2008il at $\sim$3 days after explosion (red) and SN~2007od at 6 days after explosion (green). Broadband SDSS filters are shown as dashed lines.}
\label{spectemp}
\end{figure}

We show in figure~\ref{spectemp} a comparison between the spectral template by Nugent at maximum that we use in this analysis \citep{Nugent02} with early observed spectra of nearby SNe~II \citep{Gutierrez14}. Effective wavelengths calculated with the observed spectra result in a wavelength shift of 24, 20, 64 and 2 \AA\, for $ugri$ with respect to the template.

\begin{figure}%[htbp]
\centering
\includegraphics[width=1.0\linewidth]{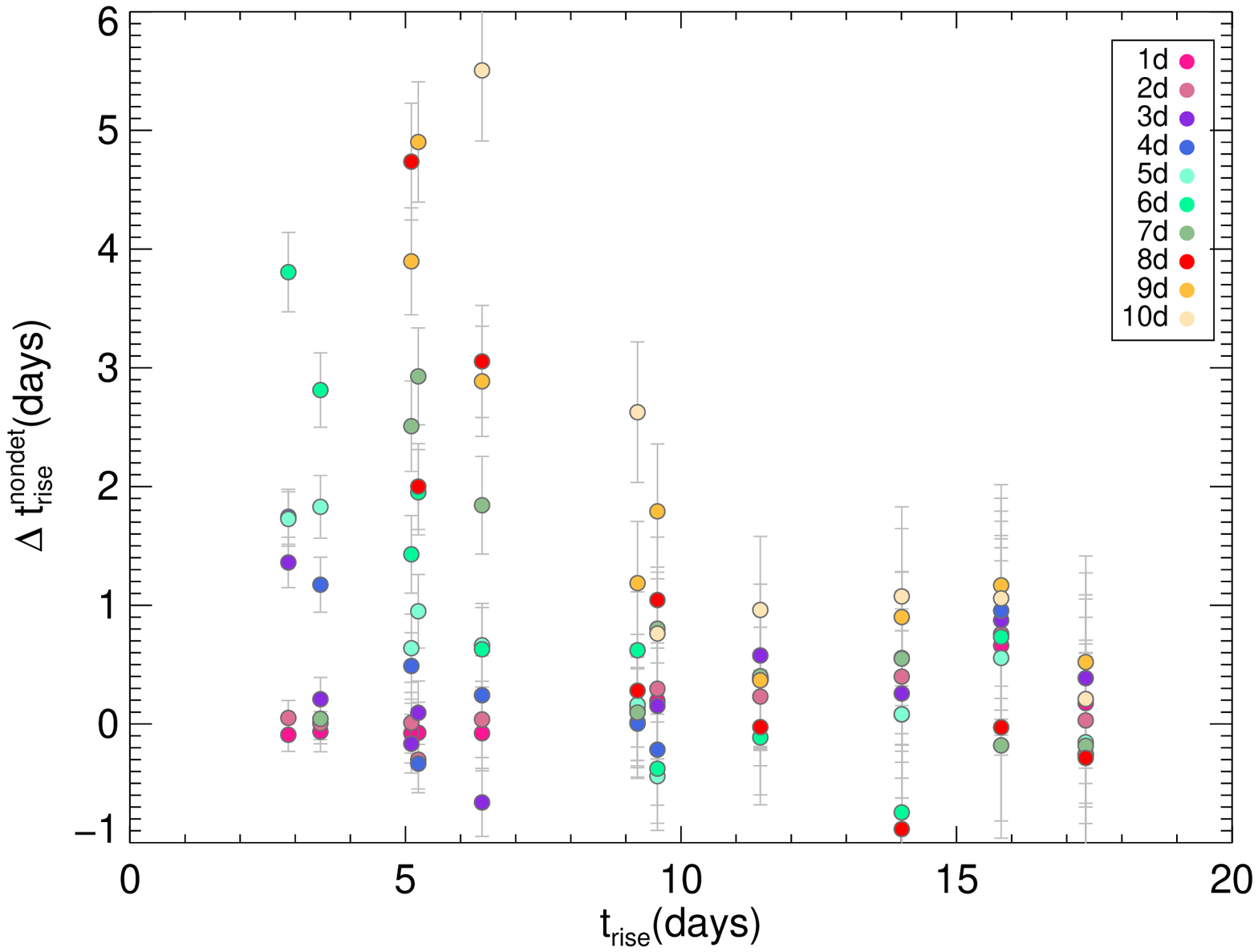}
\caption{Simulated change in rise-time, $\Delta t_{\mathrm{rise}}^{\mathrm{nondet}}$, measured with an initial date obtained from the last non-detection and first detection, as a function of rise-time, $t_{\mathrm{rise}}$, due to varying cadence in the survey. The various colors indicate cadences between 1 and 10 days. The error bars are the standard deviation divided by $N$, where $N=20$ is the number of simulations randomly changing the epoch of the first non-detection point.}
\label{simu-cadence}
\end{figure}

\subsection{Cadence}\label{cadence}

 As shown in \S~\ref{dist}, the SDSS-SN and SNLS have different cadences. This large difference leads to important systematic biases when measuring the rise. To investigate this effect further we generate mock light-curves of different rise-times based on the model of Bazin (eq~\ref{eq_baz}) by varying the cadence between 1 and 10 days. We then calculate the rise-times for these light-curves in the ways mentioned in section~\ref{analysis}. The rise-times based on initial dates from the power-law fits are generally accurate, but since we require data in the rise, there is a lack of objects with short rise-times when the cadence is high. This explains the missing fraction of SNe with short rise durations in the SNLS distribution. If, instead, we use non-detections to estimate the initial day of the rise, this effect is only partially mitigated (figure~\ref{simu-cadence}) since the end point of the rise also depends on the cadence. Moreover, the rise-times in this case are over-estimated, particularly for short rise-times. We conlcude that this method does not represent an improvement over the power-law fits. The final rise-times we calculate for the SDSS-SN are always shorter than for the SNLS, independently of the way we measure them. This difference persists even at longer wavelenghts, demonstrating that the distribution of rise-times in red bands also extends to values lower than $\sim10$ days. In fact, the median rise-time at 8200\AA\, for the golden sample of the SDSS-SN is $10.1^{+3.6}_{-2.3}$ days. This systematic effect opens up the possibility of a fraction of SNe~II with even shorter rise-times than the lowest cadences of the SDSS-SN of 2-3 days.

\subsection{Differences in starting date}
\begin{figure}%[htbp]
\centering
\includegraphics[width=1\linewidth]{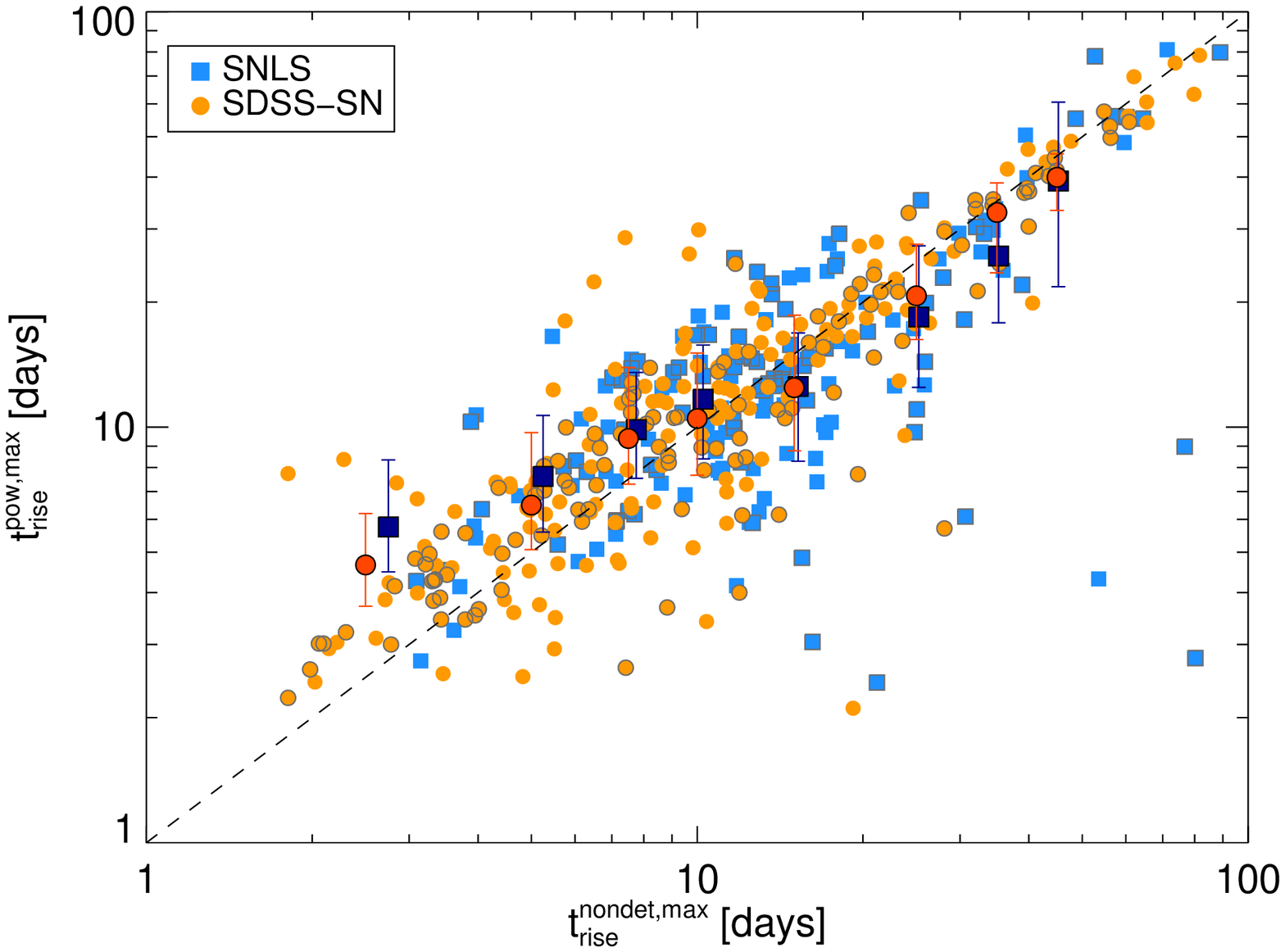}
\caption{Comparison of the effect of different starting dates used in the rise-time estimation: $t_{\mathrm{rise}}^{\mathrm{pow,max}}$ based on power-law fits versus $t_{\mathrm{rise}}^{\mathrm{nondet,max}}$ based on non-detections for the samples of SDSS-SN (orange circles) and SNLS (blue squares). Symbols with grey contours represent the golden samples and larger symbols are the median for both surveys. The dashed line indicate a 1:1 relation.}
\label{trise-exprise}
\end{figure}

We compare in Figure~\ref{trise-exprise} the difference between rise-times measured using a starting date from the power-law fit (eq.~\ref{eq_pow}), $t_{\mathrm{rise}}^{\mathrm{pow,max}}$, and with a starting date from the midpoint of the last non-detection and first detection, $t_{\mathrm{rise}}^{\mathrm{nondet,max}}$. In both cases, the endpoint of the rise is the maximum from eq.~\ref{eq_baz}, so that all differences are due to the starting date. In general, we find good agreement except at short rise-times where the power-law measurements are larger on average by 1-2 days for both surveys. We attribute this to the fact that fewer points are used in the fit, in most cases only one given the cadence of our surveys, and the first detection in the rise is closer to the time of half the flux at maximum, i.e. the limit of the time range used in the fit. This pushes the explosion date to later times. The few outliers are errors in the starting dates based on non-detections, for which fluctuations in the noise make actual non-detections sometimes inconsistent with zero in one particular band. This leads to much earlier starting dates and much longer rise-times. Given this systematic, we prefer using starting dates based on power-law fits. If the shortest measurements are slightly biased towards longer rise-times, the main conclusions of this paper are only further strengthened.

\begin{figure}%[htbp]
\centering
\includegraphics[width=1\linewidth]{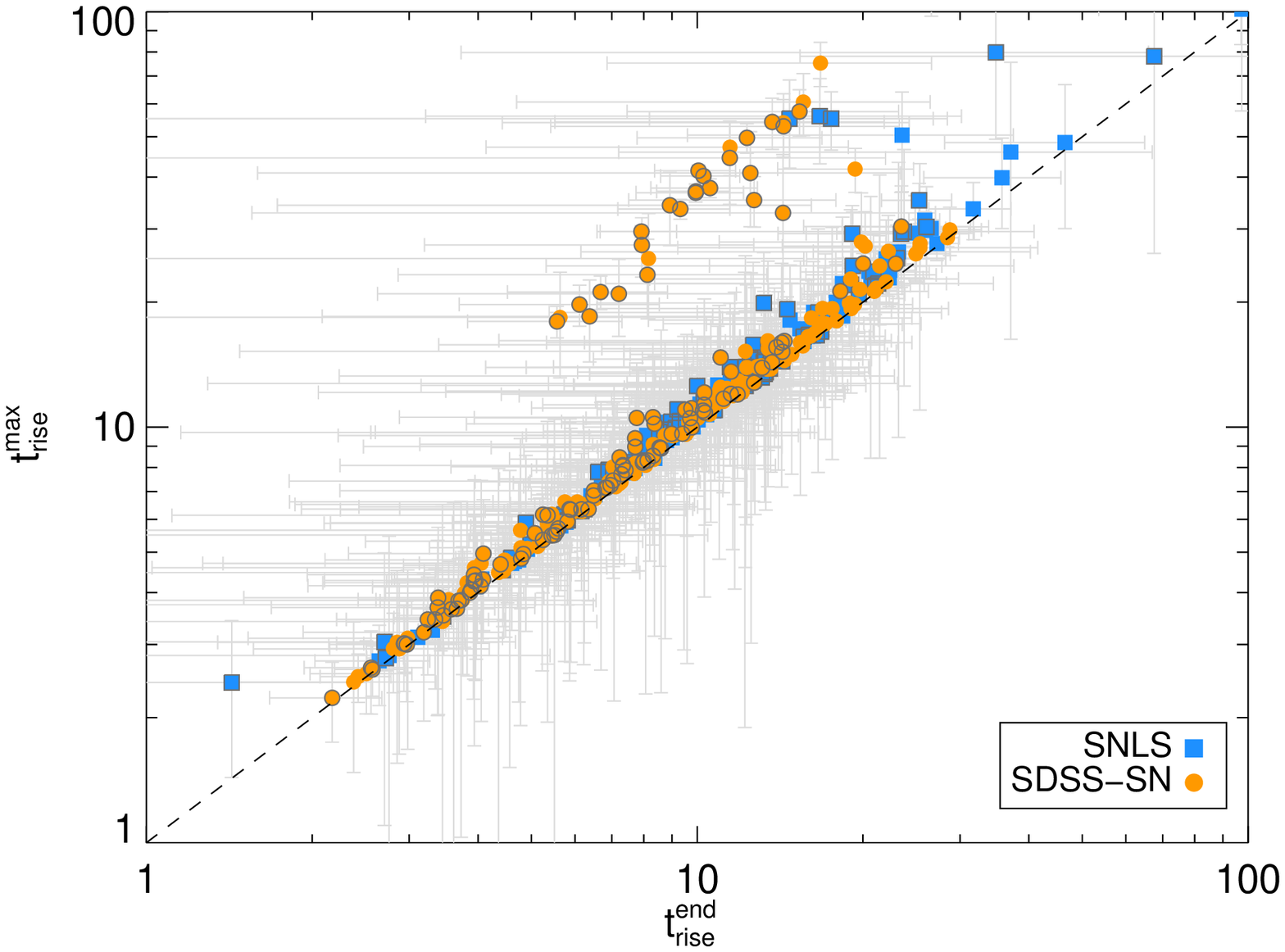}
\caption{Comparison of two rise-times measures, \rise versus \fitrise, for the samples of SDSS-SN (orange circles) and SNLS (blue squares). Symbols with contours represent the golden samples.}
\label{fitrise-exprise}
\end{figure}

\subsection{Differences in end date}
Here we compare the differences between rise-times measured using the date of maximum, \rise, and using the end date, \fitrise, but the same starting date. Figure~\ref{fitrise-exprise} shows that they are remarkably similar since they are both calculated with the same fit of Bazin (eq.~\ref{eq_baz}). However, one can see that for some light-curves at longer wavelengths, the maximum occurs during a rising plateau and is therefore quite different from the endpoint of the rise based on when the derivative becomes lower than 0.01mag/day. We thus use this date instead of the maximum in all Figures as a more characteristic end date for the rise but we quote the rise-times for the maximum in Table~\ref{table-rise}.

Our main findings of short rise-times leading to smaller radii through analytical models of NS10 and RW11, and hydrodynamical models of T09, still hold even after putting the extremes of these systematics together. We cannot account for a systematic offset of 7-8 days to recover typical RSG radii of $\sim800$\rsun\, in the analytical models.

\section{Spectra of long rise SNe}\label{longrise-spec}
\begin{figure*}%[htbp]
\centering
\includegraphics[width=0.49\linewidth]{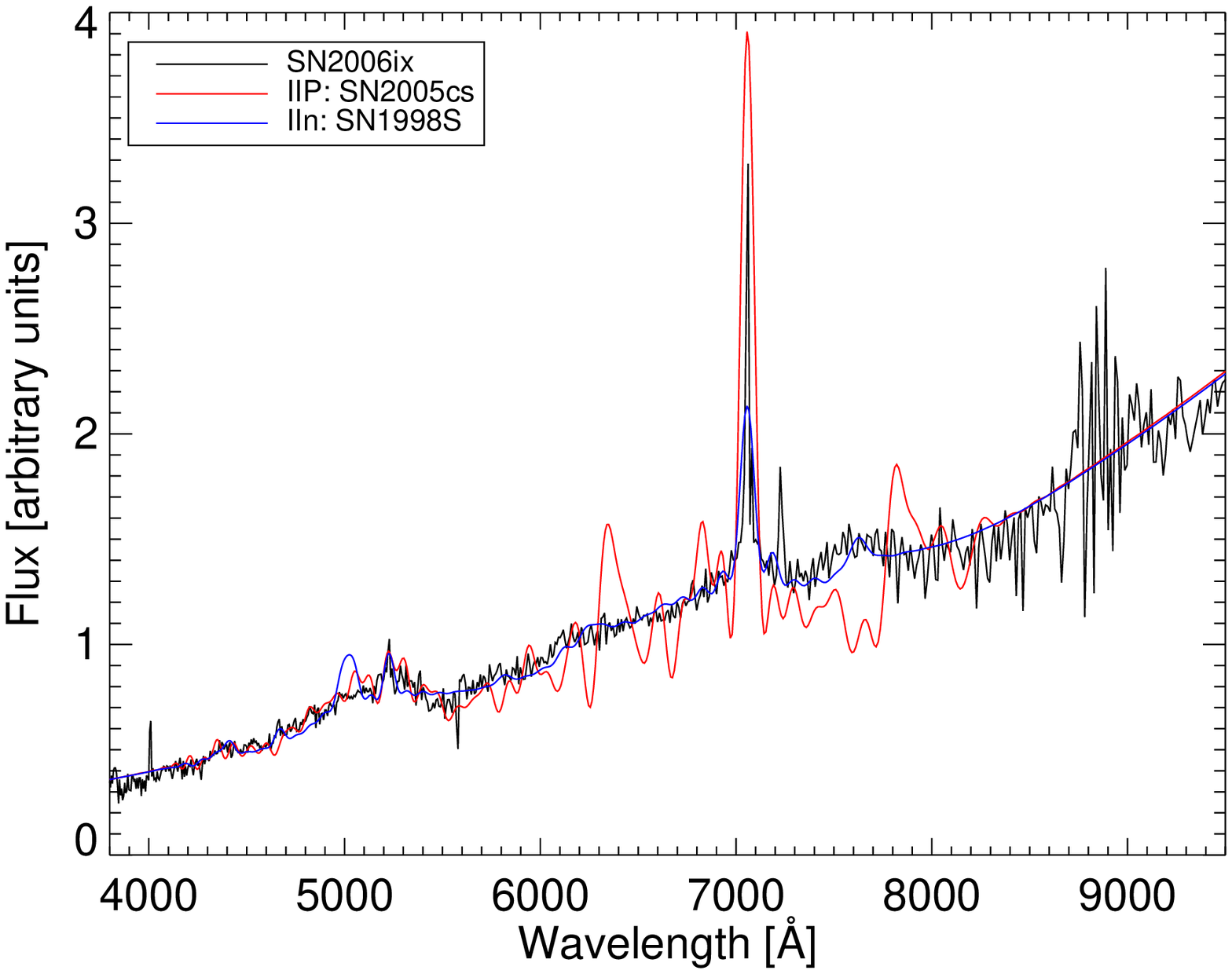}
\includegraphics[width=0.49\linewidth]{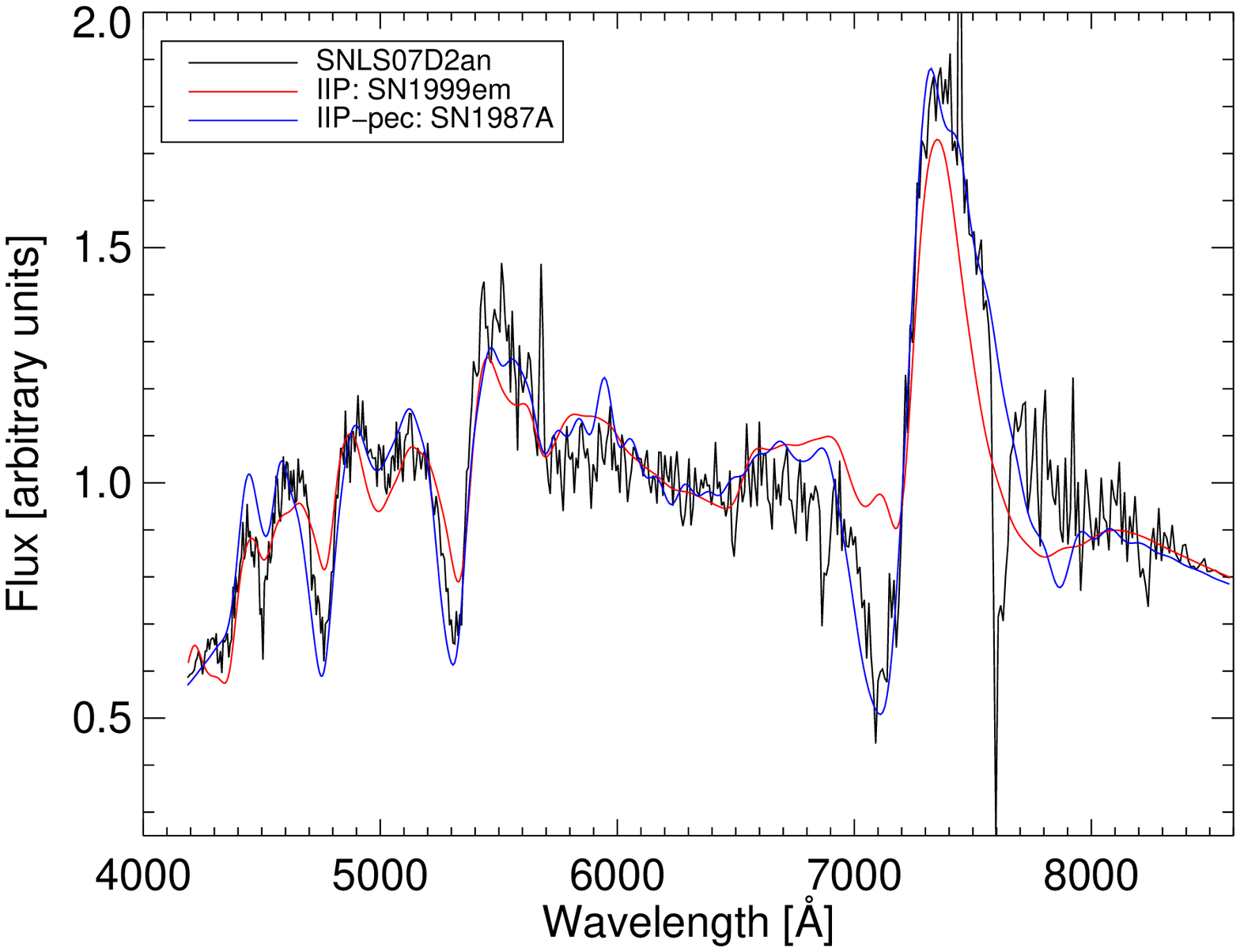}
\caption{Observed spectra of two long rise SNe initially identified as SNe~II from the post-maximum shallow slopes in arbitrary flux units. SN~2006ix (left) is compared to the best SN~II match from SNID as well as the best SN~IIn match. SNLS-07D2an (right) is compared to the best SN~II match from SNID as well as the best SN~1987A-like match.}
\label{spec}
\end{figure*}

Here we present the spectra of the two long rise SNe of section~\ref{longrise} as a reprentative case of this set of objects. The observed spectra in observer-frame compared to some SNID templates are shown in figure~\ref{spec}. For SN~2006ix the better match to SN~IIn is evident. SNLS-07D2an is an interesting SN that has very long plateaus in all bands which is unexpected for typical SN~1987A objects. The long rise-time however and the better spectral match make this a remarkable unique peculiar object.

\bsp

\label{lastpage}

\end{document}